\documentclass[prb,preprint]{revtex4}%

\usepackage{graphicx}
\usepackage{amsmath}
\usepackage{ulem}
\usepackage{color}

\newcommand{\be}{\begin{equation}}
\newcommand{\ee}{\end{equation}}
\newcommand{\bea}{\begin{eqnarray*}}
\newcommand{\eea}{\end{eqnarray*}}
\newcommand{\bean}{\begin{eqnarray}}
\newcommand{\eean}{\end{eqnarray}}

\begin{document}

\draft
\title
{\bf Thermoelectric Properties of a Semiconductor Quantum Dot
Chain Connected to Metallic Electrodes}

\author{David M.-T. Kuo$^{1,\dagger}$ and Yia-Chung Chang$^{2,*}$ }

\address{$^{1}$Department of Electrical Engineering, National Central
University, Chungli, 320 Taiwan} 

\address{$^{2}$Research Center for Applied Sciences, Academic
Sinica, Taipei, 115 Taiwan} 


\begin{abstract}
The thermoelectric properties of a semiconduct quantum dot chain
(SQDC) connected to metallic electrodes are theoretically
investigated in the Coulomb blockade regime. An extended Hubbard
model is employed to simulate the SQDC system consisted of
{\color{blue}N=2,3,4, and 5} quantum dots (QDs). The charge and heat
currents are calculated in the framework of Keldysh Green's function
technique. We obtained a closed-form Landauer expression for the
transmission coefficient of the SQDC system with arbitrary  number
of QDs by using the method beyond mean-field theory. The electrical
conductance ($G_e$), Seebeck coefficient (S), thermal conductance,
and figure of merit (ZT) are numerically calculated and analyzed in
the linear response regime. When thermal conductance is dominated by
phonon carriers, the optimization of ZT is determined by the power
factor ($pF=S^2G_e$). We find that the optimization of ZT value
favors the following conditions:(1) QDs with low energy level
fluctuations, (2) QD energy levels lie above the Fermi level of
electrodes, (3) $\Gamma < t_c \ll U_0$, where $t_c$, $U_0$, and
$\Gamma$ are electron interdot hopping strength, on-site electron
Coulomb interaction, and tunneling rate, respectively, and (4)
$\Gamma_L=\Gamma_R$ with $\Gamma_L+\Gamma_R$ kept constant, where
$\Gamma_L (\Gamma_R)$ is the left (right) tunneling rate. It is
predicted that high ZT values can be achieved by tailoring
above conditions.

\vspace{3cm}


\end{abstract}

\maketitle


\section{Introduction}
Recently, considerable studies have been devoted to seeking
efficient thermoelectric materials with the figure of merit ($ZT$)
larger than 3 because there exist potential applications of solid
state thermal devices such as coolers and power generators.$^{1-7)}$
The optimization of ($ZT=S^2G_eT/\kappa$) depends on the
thermoelectric response functions; electrical conductance ($G_e$),
Seebeck coefficient (S), and thermal conductance ($\kappa$). T is
the equilibrium temperature. These thermoelectrical response
functions are usually related to one another. Mechanisms leading to
the enhancement of power factor ($pF=S^2G_e$) would also enhance the
thermal conductance. Consequently, it is difficult to find ZT above
1 in conventional bulk materials.$^{1)}$ Nanotechnology development
provides a possible means to achieve highly efficient thermoelectric
materials. Recently, it has been demonstrated that ZT's of
nanostructure composites can reach impressive values (larger than
one).$^{8)}$ In particular, quantum dot superlattice (QDSL)
nanowires exhibit an interesting thermoelectric property in that the
power factor and thermal conductance become independent
thermoelectric variables.$^{8)}$ Based on this property, one can
increase the power factor and decrease the thermal conductance
simultaneously to optimize ZT. A ZT value close to 2 in PbSe/PbTe
QDSL system was reported.$^{6)}$ The reduction of thermal
conductance of QDSL was attributed to the increase of phonon
scattering rates, which results from phonon scattering from quantum
dot (QD) interface states.$^{1-2)}$ The thermoelectric properties of
QDSL have been theoretically studied by solving the Boltzmann
equation in reference [9].

{\color{blue}Thermoelectric properties of a single molecule and
semiconductor quantum dots (QDs) embedded in a matrix connected to
metallic electrodes were studied by several efforts.$^{10-14)}$ Very
impressive ZT values (larger than 3) were reported in previous
theoretical studies when only the electron ballistic transport was
considered.$^{10-14)}$ However, in realistic QD junctions for
thermoelectric application, one needs to consider a large number of
serially coupled QDs, otherwise it is not easy to maintain a large
temperature difference across the QD junction, which was pointed out
to be crucial in the implementation of high-efficiency
thermoelectric devices.$^{2)}$ Here we propose a tunneling system as
illustrated in Fig. 1 with a thick slab sandwiched between two
metallic leads. The slab consists of arrays of quantum wires made of
insulating material (such as SiO$_2$) surrounded by vacuum.
Temperature gradient can be easily established, since the slab has
low thermal conductivity. The quantum wires are then filled with
semiconductor QDs. For example, it is known that Si QDs embedded in
SiO$_2$ can be fabricated. It is expected that the electrical
conductivity of this junction system may be enhanced via electron
hopping through the QD chain. The conductivity of the junction
system can be tuned by the density of QDs. }

Here we consider nanoscale semiconductor QDs, in which the energy
level separations are much larger than their on-site Coulomb
interactions and thermal energies. Thus, only one energy level for
each quantum dot needs to be considered. An $N$-level Anderson model
is employed to simulate the system as shown Fig. 1.$^{15)}$
Increasing the number of QDs would change the DOS of the junction
system from atomic limit to the band limit. When N is infinite, it
can be regarded as model system to clarify fundamental physics of
one-dimensional strongly-correlated systems, which is one of the
most challenging problems in condensate matter physics.$^{15)}$ For
a semiconductor quantum dot chain (SQDC) consisted of finite number
of coupled QDs, its transport properties are still very complicated
as a result of electron Coulomb interactions and interdot hopping
effects. Consequently, it is tedious to theoretically obtain the
best ZT value. This study has derived an analytical formula of
charge and heat currents to avoid such a difficulty. The main goal
of this article attempts to reveal the effects of electron Coulomb
interaction, interdot hopping strength, QD size fluctuation, and QD
number on the ZT optimization of insulator junction systems.




\section{Formalism}

The Hamiltonian of $N$ coupled QDs connected to metallic electrodes
can be described by the combination of extended Hubbard model and
Anderson model $H=H_0+H_{QD}$:

\begin{eqnarray}
H_0& = &\sum_{k,\sigma} \epsilon_k
a^{\dagger}_{k,\sigma}a_{k,\sigma}+ \sum_{k,\sigma} \epsilon_k
b^{\dagger}_{k,\sigma}b_{k,\sigma}\\ \nonumber &+&\sum_{k,\sigma}
V_{k,1}d^{\dagger}_{1,\sigma}a_{k,\sigma}
+\sum_{k,\sigma}V_{k,N}d^{\dagger}_{N,\sigma}b_{k,\sigma}+c.c
\end{eqnarray}
where the first two terms describe the free electron gas of left and
right metallic electrodes. $a^{\dagger}_{k,\sigma}$
($b^{\dagger}_{k,\sigma}$) creates  an electron of momentum $k$ and
spin $\sigma$ with energy $\epsilon_k$ in the left (right) metallic
electrode. $V_{k,\ell}$ ($\ell=1,N$) describes the coupling between
the metallic electrode and the first ($N$-th) QD.
$d^{\dagger}_{\ell,\sigma}$ ($d_{\ell,\sigma}$) creates (destroys)
an electron in the $\ell$-th dot.

\begin{small}
\begin{eqnarray}
H_{QD}&=& \sum_{\ell,\sigma} E_{\ell} n_{\ell,\sigma}+
\sum_{\ell} U_{\ell} n_{\ell,\sigma} n_{\ell,\bar\sigma}\\
\nonumber &+&\frac{1}{2}\sum_{\ell \neq j,\sigma,\sigma'}
U_{\ell,j}n_{\ell,\sigma}n_{j,\sigma'} +\sum_{\ell \neq
j,\sigma}t_{\ell,j} d^{\dagger}_{\ell,\sigma} d_{j,\sigma},
\end{eqnarray}
\end{small}
where { $E_{\ell}$} is the spin-independent QD energy level, and
$n_{\ell,\sigma}=d^{\dagger}_{\ell,\sigma}d_{\ell,\sigma}$.
Notations $U_{\ell}$ and $U_{\ell,j}$ describe the intradot and
interdot Coulomb interactions, respectively. $t_{\ell,j}$ describes
the electron interdot hopping. Note that the nearest neighbor
interdot hopping and interdot Coulomb interaction are taken into
account in Eq. (2). For simplicity, we assume that electrons
transport in a ballistic process.

Using the Keldysh-Green's function technique,$^{16)}$ the charge
current leaving the left and right electrodes in the steady state
can be expressed by
\begin{equation}
J_L=\frac{-e}{h}\int d\epsilon \Gamma_L
[2f_L(\epsilon)ImG^r_{1,\sigma}(\epsilon)
-iG^{<}_{1,\sigma}(\epsilon)]
\end{equation}
and
\begin{equation}
J_R=\frac{-e}{h}\int d\epsilon \Gamma_R
[2f_R(\epsilon)ImG^r_{N,\sigma}(\epsilon)-iG^{<}_{N,\sigma}(\epsilon)],
\end{equation}
where $\Gamma_{L}=\Gamma_{1}$ and $\Gamma_{R}=\Gamma_{N}$ denote,
respectively, the tunneling rates of the left electrode to the first
QD and the right electrode to the $N$th QD, which are assumed to be
energy and bias independent for simplicity.
$f_{L(R)}(\epsilon)=1/[e^{(\epsilon-\mu_{L(R)})/k_BT_{L(R)}}+1]$
denotes the Fermi distribution function for the left (right)
electrode. The chemical potential difference is given by
$\mu_L-\mu_R=e\Delta V$. $T_{L(R)}$ denotes the equilibrium
temperature of the left (right) electrode. $e$ and $h$ denote the
electron charge and Planck's constant, respectively. Obviously,
tunneling currents are determined by the on-site retarded Green's
function ($G^{r}_{1(N),\sigma}(\epsilon)$) and lesser Green's
function ($G^{<}_{1(N),\sigma}(\epsilon)$). It is not trivial to
solve N coupled QDs with electron Coulomb interactions ($U_{\ell}$
and $U_{\ell,j}$) and electron interdot hopping ($t_{\ell,j}=t_c$).
Conventional Hartree-Fork mean field theory can not resolve the
detailed quantum pathes.  Our previous approach beyond the
mean-field approximation can obtain all Green functions in the
closed form solutions for N=2 in the limit of $t_c/U_{\ell} \ll
1$.$^{13,14)}$ Based on the same approach, we have demonstrated that
the charge and heat currents of SQDC can be expressed as

\begin{eqnarray}
J&=&\frac{2e}{h}\int d\epsilon {\cal T}(\epsilon)
[f_L(\epsilon)-f_R(\epsilon)],\\ Q&=& \frac{2}{h}\int d\epsilon
{\cal T}(\epsilon)(\epsilon-E_F-e\Delta V)
[f_L(\epsilon)-f_R(\epsilon)],
\end{eqnarray}
where ${\cal T}(\epsilon)\equiv ({\cal T}_{1,N}(\epsilon) +{\cal
T}_{N,1}(\epsilon))/2$ is the transmission coefficient, $E_F$ is the
Fermi energy of electrodes. ${\cal T}_{\ell,j}(\epsilon)$ denotes
the transmission function, which can be expressed in terms of the
on-site retarded Green's functions, even though ${\cal
T}_{\ell,j}(\epsilon)$ should be calculated by the on-site retarded
and lesser Green's functions. The transmission function in the weak
interdot hopping limit ($t_{c} \ll U_{\ell}$) has the following
form,
\begin{equation}
{\cal T}_{1,N}(\epsilon)=-2\sum^{2\times
4^{N-1}}_{m=1}\frac{\Gamma_{1}(\epsilon)
\Gamma^{m}_{1,N}(\epsilon)}{\Gamma_{1}(\epsilon)+\Gamma^{m}_{1,N}(\epsilon)}
\mbox{Im}G^r_{1,m}(\epsilon),
\end{equation}
where Im means taking the imaginary part of the function that
follows, and
\begin{equation}
G^r_{1,m}(\epsilon)=p_{1,m}/(\mu_{1}-\Pi_{1,m}-\Sigma^m_{1,N}),
\end{equation}
where $\mu_{\ell}=\epsilon-E_{\ell}+i\Gamma_{\ell}/2$. Note that
$\Gamma_{\ell}=0$ when $\ell \neq 1(N)$. $\Pi_{1,m}$ denotes the sum
of Coulomb energies arising from other electron present in the first
QD and its neighborhood, and
$\Gamma^m_{1,N}(\epsilon)=-2Im\Sigma^m_{1,N}(\epsilon)$, where
$\Sigma^m_{1,N}$ denotes the self energy resulting from electron
hopping from QD 1 to QD $N$ through channel $m$. For electron with
spin $\sigma$ tunneling from the left electrode into level 1 of $N$
serially coupled QD and exit to the right electrode, we have
$2\times 4^{N-1}$ quantum pathes (or channels), since level 1 can be
either empty or singly occupied (with spin $-\sigma$) and all other
levels can be empty, singly occupied (with spin up or down), and
doubly occupied. For the $N=2$ case, the explicit expressions of
probability weights and self energies have been worked out in our
previous paper.$^{14)}$ Here, we generalize the expressions to a
SQDC system with arbitrary number of QDs, which is correct in the
limit $t_{c}/U_{\ell} \ll 1$. We found that for the $N$-QD system,
the probability factors $p_{1,m}$ are determined by the following
relation \be(\bar a_1+\bar b_{1\bar\sigma})\prod_{\ell=2}^N
(a_\ell+b_{\ell\sigma}+
b_{\ell\bar\sigma}+c_\ell)=\sum_{m=1}^{2\times 4^{N-1}} p_{1,m},\ee
where $\bar a_{\ell}=1-N_{\ell,\bar\sigma}$, $\bar
b_{\ell,\bar\sigma}=N_{\ell,\bar\sigma}$,
$a_{\ell}=1-N_{\ell,\bar\sigma}-N_{\ell,\sigma}+c_{\ell}$,
$b_{\ell,\bar\sigma}=N_{\ell,\bar\sigma}-c_{\ell}$,
$b_{\ell,\sigma}=N_{\ell,\sigma}-c_{\ell}$. $a_{\ell}$,
$b_{\ell,\bar\sigma}$, $b_{\ell,\sigma}$, and $c_{\ell}$ describe
the probability factor for the $\ell$-th QD with no electron, one
electron of spin $\bar\sigma$, one electron of spin $\sigma$, and
two electrons, respectively. $N_{\ell,\sigma}$,
$N_{\ell,\bar\sigma}$, and  $c_{\ell}=\langle n_{\ell,\bar\sigma}
n_{\ell,\bar\sigma} \rangle$ denote the thermally averaged
one-particle occupation numbers for spin $\sigma$ and $\bar\sigma$
and two-particle correlation functions. We note that the sum of Eq.
(9) equals one, which indicates probability conservation.
{\color{blue} $\Sigma^m_{1,N}$ in Eq. (8) is given by

\begin{equation}
\Sigma^m_{1,N}=\frac{t^2_{1,2}}{\mu_2-\Pi_{2,m}-\frac{t^2_{2,3}}{\mu_3-\Pi_{3,m}
\cdots -\frac{t^2_{N-1,N}}{\mu_N-\Pi_{N,m} }, }}
\end{equation}

where $\Pi_{\ell m}$ denotes the sum of Coulomb energies due to
interaction of an electron entering the $\ell$-th QD with the other
electrons present in the SQDC in configuration $m$. In the absence
of electron Coulomb interactions $\Sigma^m_{1,N}$ reduces to the
continued fraction results for a un-correlated linear chain model,
which can be obtained by the recursion method. For N=3, the detailed
expression of $\Sigma^m_{1,N}$ is provided in the appendix. Our
results are derived from solving the equation of motion for N weakly
coupled QDs with strong correlation according to similar procedures
described in the Appendix of Ref. 13. Our results should not be
confused with the results obtained from the so called Hubbarb-I
approximation$^{17)}$, which is essentially a mean-field approach.
In our approach, the charging energies $\Pi_{\ell,m}$ appearing in
Eq. (10) are associated with integer charges, which do not depend on
the occupation number and two-particle correlation function. If one
employs the Hubbard-I approximation to truncate the high order Green
functions, the occupation number will appear in the denominator of
Eq. (8).$^{17)}$ The $\Pi_{N,m}$ with integral charges is the
manifestation of strong correlation, which is obtained by a
many-body theory employed to treat high order Green functions
arising from electron Coulomb interactions beyond the Hubbard-I
approximation.$^{13)}$}

The average occupation numbers $N_{\ell,\sigma}=N_{\ell,\bar\sigma}$
and  $c_{\ell}$ are determined by solving the on-site lesser Green's
functions $iG^{<}_{1,\sigma}(\epsilon)$,$^{13,14)}$ which take the
following form
\begin{equation}
iG^{<}_{1,\sigma}(\epsilon)=2\sum_{m}\frac{\Gamma_1(\epsilon)
f_L+\Gamma^m_{1,N}(\epsilon)f_R}{
\Gamma_1(\epsilon)+\Gamma^m_{1,N}(\epsilon)}ImG^r_{1,m}(\epsilon).
\end{equation}

Thus, we have
\begin{equation}
N_1=-\int \frac{d\epsilon}{\pi} \sum_{m}\frac{\Gamma_1(\epsilon)
f_L+\Gamma^m_{1,N}(\epsilon)f_R}{
\Gamma_1(\epsilon)+\Gamma^m_{1,N}(\epsilon)}ImG^r_{1,m}(\epsilon),
\end{equation}
and
\begin{equation}
c_1=-\int \frac{d\epsilon}{\pi} \sum_{m=4^{N-1}+1}^{2\times
4^{N-1}}\frac{\Gamma_1(\epsilon) f_L+\Gamma^m_{1,N}(\epsilon)f_R}{
\Gamma_1(\epsilon)+\Gamma^m_{1,N}(\epsilon)}ImG^r_{1,m}(\epsilon),
\end{equation}
where $\sum_{m=4^{N-1}+1}^{2\times 4^{N-1}}$ denotes a sum over
configurations obtained by the product \be
N_{1\bar\sigma}\prod_{\ell=2}^N
(a_{\ell}+b_{\ell,\sigma}+b_{\ell,\bar\sigma}+c_{\ell})=\sum_{m=4^{N-1}+1}^{2\times
4^{N-1}} p_{1,m}. \ee To calculate the tunneling current from of Eq.
(3), we also need the retarded Green's function
$G^r_{1,\sigma}(\epsilon)$, which is given by
$G^r_{1,\sigma}(\epsilon)=\sum_{m} G^r_{1,m}(\epsilon)$.

$N_{N,\sigma}=N_{N,\bar\sigma}=N_N$ and $c_N$ have the same forms as
the above equations with the indices 1 and $N$ exchanged.
$G^r_{N,m}(\epsilon)$ is obtained from $G^r_{1,m}(\epsilon)$ by
reversing the roles of QDs 1 and $N$. Namely,
\begin{equation}
G^r_{N,m}(\epsilon)=p_{N,m}/(\mu_{N}-\Pi_{N,m}-\Sigma^m_{N,1}),
\end{equation}
\be
\Sigma^m_{N,1}=\frac{t^2_{N,N-1}}{\mu_{N-1}-\Pi_{N-1,m}-\frac{t^2_{N-1,N-2}}{\mu_{N-2}-\Pi_{N-2,m}
\cdots -\frac{t^2_{2,1}}{\mu_1-\Pi_{1,m} }, }} \ee and $p_{N,m}$ are
determined by \be (\bar a_N+\bar
b_{N\bar\sigma})\prod_{\ell=1}^{N-1}
(a_\ell+b_{\ell\sigma}+b_{\ell,\bar\sigma}+c_{\ell})=\sum_{m=1}^{2\times
4^{N-1}} p_{N,m}. \ee

For QDs not in direct contact with leads (labeled by $\ell=2,N-1$),
we have
\begin{equation}
N_{\ell}=-\int \frac{d\epsilon}{\pi} \sum_{m=1}^{2\times
4^{N-1}}\frac{\Gamma^{m}_{\ell,1}(\epsilon)
f_L+\Gamma^m_{\ell,N}(\epsilon)f_R}{
\Gamma^{m}_{\ell,1}(\epsilon)+\Gamma^m_{\ell,N}(\epsilon)}ImG^r_{\ell,m}(\epsilon),
\end{equation}
and
\begin{equation}
c_\ell=-\int \frac{d\epsilon}{\pi} \sum_{m=4^{N-1}+1}^{2\times
4^{N-1}}\frac{\Gamma^{m}_{\ell,1}(\epsilon)
f_L+\Gamma^m_{\ell,N}(\epsilon)f_R}{
\Gamma^{m}_{\ell,1}(\epsilon)+\Gamma^m_{\ell,N}(\epsilon)}ImG^r_{\ell,m}(\epsilon).
\end{equation}
$\Gamma^m_{\ell,1}(\epsilon)=-2Im\Sigma^m_{\ell,1}$ and
$\Gamma^m_{\ell,N}(\epsilon)=-2Im\Sigma^m_{\ell,N}$ are the
effective tunneling rates for electrons from the $\ell$-th QD to the
left and right electrodes, respectively. The retarded Green's
function $G^r_{\ell,m}(\epsilon)$ is given by

\be G^r_{\ell,m}(\epsilon)=
\frac{p_{\ell,m}}{\mu_\ell-\Pi_{\ell,m}-\Sigma^{m}_{\ell,1}-\Sigma^{m}_{\ell,N}
}, \ee where the probability factors, $p_{\ell,m}$  are determined
by Eq. (9)  with the indies 1 and $\ell$ interchanged. The self
energies $\Sigma^m_{\ell,1}$ and $\Sigma^m_{\ell,N}$ are given by
\be
\Sigma^m_{\ell,1}=\frac{t^2_{\ell,\ell-1}}{\mu_{\ell-1}-\Pi_{\ell-1,m}-\frac{t^2_{\ell-1,\ell-2}}{\mu_{\ell-2}-\Pi_{N-2,m}
\cdots -\frac{t^2_{2,1}}{\mu_1-\Pi_{1,m} } }} \ee and \be
\Sigma^m_{\ell,N}=\frac{t^2_{\ell,\ell+1}}{\mu_{\ell+1}-\Pi_{\ell+1,m}-\frac{t^2_{\ell+1,\ell+2}}
{\mu_{\ell+2}-\Pi_{\ell+2,m} \cdots
-\frac{t^2_{N-1,N}}{\mu_N-\Pi_{N,m} }. }} \ee

The explicit expressions of $p_{\ell,m}$ and $\Pi_{\ell,m}$ for all
levels and all configurations of the N=3 case are given in the
appendix. As an example, for a five-dot SQDC with configuration
described by $p_{1,m}=a_1 b_{2\sigma}a_3 b_{4,\bar\sigma} c_5$, we
have
\[
G^r_{1,m}(\epsilon)= \]
\begin{equation}
\frac{(1-N_{1,\bar\sigma}) b_{2\sigma}a_3 b_{4,\bar\sigma} c_5}
{\mu_1-U_{1,2}-\frac{t^2_{1,2}}{\mu_2-U_{1,2}-\frac{t^2_{2,3}}{\mu_3-U_{3,4}-\frac{t^2_{3,4}}{\mu_4-U_4-2U_{4,5}
-\frac{t^2_{4,5}}{\mu_5-U_5 -2U_{4,5}}.} }}}
\end{equation}

In the linear response regime, Eqs. (5) and (6) can be rewritten as
\begin{eqnarray}
J&=&{\cal L}_{11} \frac{\Delta V}{T}+{\cal L}_{12} \frac{\Delta
T}{T^2}\\Q&=&{\cal L}_{21} \frac{\Delta V}{T}+{\cal L}_{22}
\frac{\Delta T}{T^2},
\end{eqnarray}
where there are two sources of driving force to yield the charge and
heat currents. $\Delta T=T_L-T_R$ is the temperature difference
across the junction. Thermoelectric response functions ${\cal
L}_{11}$, ${\cal L}_{12}$, ${\cal L}_{21}$, and ${\cal L}_{22}$ are
evaluated by

\begin{equation}
{\cal L}_{11}=\frac{2e^2T}{h} \int d\epsilon {\cal T}(\epsilon)
(\frac{\partial f(\epsilon)}{\partial E_F})_T,
\end{equation}
\begin{equation}
{\cal L}_{12}=\frac{2eT^2}{h} \int d\epsilon {\cal T}(\epsilon)
(\frac{\partial f(\epsilon)}{\partial T})_{E_F},
\end{equation}

\begin{equation}
{\cal L}_{21}=\frac{2eT}{h} \int d\epsilon {\cal
T}(\epsilon)(\epsilon-E_F) (\frac{\partial f(\epsilon)}{\partial
E_F})_T,
\end{equation}
and
\begin{equation}
{\cal L}_{22}=\frac{2T^2}{h} \int d\epsilon {\cal T}(\epsilon)
(\epsilon-E_F)(\frac{\partial f(\epsilon)}{\partial T})_{E_F}.
\end{equation}
Here ${\cal T}(\epsilon)$ and
$f(\epsilon)=1/[e^{(\epsilon-E_F)/k_BT}+1]$ are evaluated under the
equilibrium condition.

If the system is in an open circuit, the electrochemical potential
will be established in response to a temperature gradient; this
electrochemical potential is known as the Seebeck voltage (Seebeck
effect). The Seebeck coefficient (amount of voltage generated per
unit temperature gradient) is defined as $S=\Delta V/\Delta T=-{\cal
L }_{12}/(T{\cal L}_{11})$. To judge whether the system is able to
generate or extract heat efficiently, we need to consider the figure
of merit,[1] which is given by
\begin{eqnarray}
ZT=\frac{S^2G_eT}{\kappa_e+\kappa_{ph}}\equiv
\frac{(ZT)_0}{1+\kappa_{ph}/\kappa_e}.
\end{eqnarray}
Here $G_e={\cal L}_{11}/T$ is the electrical conductance and
$\kappa_e= (({\cal L}_{22}/T^2)-{\cal L}_{11}S^2)$ is the electron
thermal conductance. $(ZT)_0$ represents the $ZT$ value in the
absence of phonon thermal conductance, $\kappa_{ph}$.
{\color{blue}We assume that the nanowires are surrounded by vacuum,
where phonons cannot propagate. Thus, the nanowire is the main
channel for phonons to propagate between the electrodes, and the
phonon thermal conductance of the system shown in Fig. 1 is given by
the nanowire with QDs. The phonon thermal conductance adopted,
$\kappa_{ph}=\frac{\pi^2k_B^2T}{3h}F_s$ matches very well with a
recent experiment, when considered rough silicon nanowires having
diameter smaller than 50 nm.$^{18)}$}
$\kappa_{ph,0}=\frac{\pi^2k^2_B T}{3h}$ is the universal phonon
thermal conductance arising from acoustic phonon confinement in a
nanowire,$^{19-20)}$ which was confirmed in the phonon wave
guide.$^{21)}$ The expression of $\kappa_{ph}=\kappa_{ph,0} F_s$
with $F_s=0.1$ can explain well the phonon thermal conductance of
silicon nanowire with surface states calculated by the
first-principles method.$^{19)}$ The dimensionless scattering factor
$F_s$ arises from phonon scattering with surface impurities or
surface defects of quantum dots.$^{1)}$It is possible to reduce
phonon thermal conductance by one order of magnitude when the QD
size is much smaller than the phonon mean free paths.$^{22,23)}$
Therefore, we adopt $F_s=0.01$ as a fixed parameter and assume $F_s$
is independent of the number of QDs and QD size.

\section{Results and discussion}

\subsection{Three-QD junction}
In this section, we study the thermoelectric properties of N=3 case,
which were experimentally$^{24,25)}$ and theoretically$^{26,27)}$
investigated in the nonlinear response regime to reveal the coherent
and spin-dependent behavior of carrier transport. In Ref.~[28], the
Kondo transport of triple QDs was investigated by using the
slave-boson method to remove the double occupation for each QD. This
study is restricted in the Coulomb blockade regime. So far, few
literatures have considered the spin-dependent thermoelectric
properties of three coupled QDs. In Fig. 2 we plot the electrical
conductance ($G_e$), Seebeck coefficient (S), and electron thermal
conductance ($\kappa_e$) as a function of gate voltage for various
temperatures. We adopt the following physical parameters
$U_{\ell}=U_0=60\Gamma_0$, $U_{\ell,j}=20\Gamma_0$,
$t_{\ell,j}=1\Gamma_0$, $\Gamma_1=\Gamma_3=\Gamma=\Gamma_0$,
$E_1=E_2=E_F-20\Gamma_0$, and $E_3=E_F+30\Gamma_0-eV_g$. All energy
scales are in units of $\Gamma_0$, a characteristic energy. A gate
voltage is applied to tune the energy level of $E_3$ such that the
$E_3$ level can be varied from being empty to singly occupied (see
the inset of Fig.2). The system with spin triple state as
illustrated in the inset of Fig. 2(a) is in the insulating state
when $E_3$ is far above the Fermi energy. (Note that dot 1 and dot 2
forms the spin-triplet state filter). The conductance $G_e$ reaches
a maximum at $E_3=E_F$, whose magnitude decreases as the temperature
increases. Meanwhile, $G_e$ a function of $V_g$ has a Lorentz shape,
whose FWHM is almost independent of temperature as long as
$k_BT/\Gamma_0 \ge 1$. This is referred to as the nonthermal
broadening effect, which was also observed in the case of serially
coupled quantum dots (SCQD).$^{29)}$ The behavior of Seebeck
coefficient ($S$) is illustrated in Fig. 2(b), and it is found that
$S$ vanishes when $G_e$ reaches the maximum, a result attributed to
the electron-hole symmetry. Here, holes are defined as missing
electrons in electrodes below $E_F$. The negative sign of $S$
indicates that electrons are majority carries, which diffuse to the
right electrode from the left electrode through energy levels above
$E_F$. On the other hand, holes become majority carriers when $S$
turns positive. We see that $S$ vanishes again at $k_BT=1\Gamma_0$
and $eV_g=50\Gamma_0$, where $E_3+U_J$ is lined up with $E_F$. Thus,
the measurement of $S$ can more reveal the properties of resonant
channels than that of $G_e$. The behavior of electron thermal
conductance $\kappa_e$ is similar to that of $G_e$ and also shows
the nonthermal broadening effect. Like the two-QD junction, the
nonthermal broadening effect of the three-QD junction can be used to
function as a low temperature filter.

In Fig.~3, we show thermoelectric behaviors of a system with
$E_1=E_F-20\Gamma_0$, $E_2=E_F+10\Gamma_0-eV_g$, and $E_3=E_F$, as
illustrated in the inset of Fig.~3(a). Other physical parameters are
the same as those of Fig.~2. Now the gate voltage is used to tune
the level $E_2$ such that the $E_2$ level varies from being empty to
singly occupied. Meanwhile, the $E_3$ level will be depleted when
$E_2$ is singly occupied. Although the behavior of $G_e$ shown in
Fig.~3(a) is very similar to that of Fig.~2(a), we note the FWHM of
$G_e$ in Fig.~3(a) is nearly twice as large. Furthermore, the
nonthermal broadening effect for $\kappa_e$ disappears. This
indicates that the ``effective broadening'' of energy level of dot
2, which is not directly coupled to electrodes, is different from
that of dots 1 and 3. The nonthermal broadening effect of $G_e$ is
an essential characteristic of resonant junction system. Once $k_BT$
is larger than the tunneling rates $\Gamma_1=\Gamma_3=\Gamma$, which
is the broadening of energy levels of dots 1 and 3, the broadening
of $G_e$ depends mainly on the lifetime of the resonance, and
becomes insensitive to the temperature factor
$1/\cosh^2((\epsilon-E_F)/(2k_BT))$. In addition, we find more
oscillatory peaks of $S$ in Fig.~3(b) as compared in Fig.~2(b). For
example, when $eV_g=10\Gamma_0$ we have $E_2=E_F$, which matches
with $E_3$ and $E_1+U_{12}$. This resonance has very small
probability weight and is non-observable in $G_e$, whereas it can be
measured by $S$. The highly oscillatory behavior of $S$ with respect
to $V_g$ indicates that carriers with high energies can diffuse to
the right electrode through more resonant channels, which are far
above $E_F$. This explains why the tail of $\kappa_e$ peak (near
$eV_g=20\Gamma_0$ and $eV_g=40\Gamma_0$) increases with increasing
temperature. The behavior of $\kappa_e$ can be different from that
of $G_e$.

In Fig.~4, we plot the average occupation number $N_{\ell}$, $G_e$
and $S$ as functions of the gate voltage $V_g$ (which is used to
tune $E_2$) for various temperatures for a three-QD junction system
in the spin-blockade configuration, as illustrated in the inset of
Fig.~4(b). Here, $E_1=E_F-10\Gamma_0$, $E_2=E_F+10\Gamma_0-eV_g$,
$E_3=E_F-60\Gamma_0$, and $U_{\ell,j}=10\Gamma_0$.  Other physical
parameters are the same as those of Fig.~3. This configuration was
considered in Ref. [27] within the framework of Master equation
technique for studying spin-blockade behavior of three coupled QD in
the nonlinear response regime. A spin-blockade can occur, because
the second electron appears in dot 3 must satisfy the Pauli
exclusion principle. As seen in Fig.~4(a), the $E_2$ level is tuned
from being empty to singly occupied, while the $E_3$ level remains
singly occupied even in presence of the interdot Coulomb
interactions. We observe a small bump of $G_e$ near
$eV_g=20\Gamma_0$. Although both $E_1$ and $E_2$ become resonant at
$eV_g=20\Gamma_0$, electrons in this resonance state can not tunnel
to the right electrode through $E_3$, due to the presence of
$U_{23}$ arising in dot 3. The maximum $G_e$ at $eV_g=30\Gamma_0$ is
resulting from the resonance: $E_1+U_{12}=E_2+U_{12}+U_{23}=E_3+U_3$
(described by $p_{1,10}$ of appendix), which is the Pauli spin
blockade process for the case of three-QD junction. The maximum
$G_e$ in this spin singlet state is small than the maximum of $G_e$
in spin triplet state of Fig.~2. From the results of
Figs.~(2)-(4),where quantum dots with mixture of different sizes, we
find that the maximum Seebeck coefficients are smaller than one.
This is not preferred for the purpose of enhancing ZT. Such a result
also implies that the fluctuation of QD energy levels in SQDC will
suppress ZT.

In Fig.~5, we plot the occupation numbers $N_{\ell}$, electrical
conductance $G_e$, Seebeck coefficient $S$, and electron thermal
conductance $\kappa_e$ as functions of gate voltage at
$k_BT=1\Gamma_0$. We consider identical energy levels with
$E_{\ell}=E_0=E_F+30\Gamma_0-eV_g$ and $t_{\ell,j}=6\Gamma_0$. All
QD levels are tuned by the gate voltage from far above $E_F$ to far
below $E_F$. Other physical parameters are the same as those of Fig.
2. Because $t_{\ell,j}=t_c > k_B T$ and  $t_{\ell,j}=t_c > \Gamma$,
these thermal response functions ($G_e$, S, and $\kappa_e$) display
structures yielded by the electron hopping effect. The three peaks
labeled by $V_{g1}$, $V_{g2}$, and $V_{g3}$ correspond to three
resonant channels at $E_0-\sqrt{2}~t_c$, $E_0$, and
$E_0+\sqrt{2}~t_c$, which are poles of the Green's function
$G^r_{1,m}(\epsilon)$ for channel $m=1$, in which all three QDs are
empty. The strengths of these peaks are determined by their
probability weights, $\bar a_1 a_2 a_3$. Another three peaks labeled
by $V_{g4}$, $V_{g5}$, and $V_{g6}$ result from the resonances
corresponding to channel $m=28$ with $p_{1,m}=N_{1\bar\sigma}
b_{2\sigma} c_3$, in which dots 1 and 3 are doubly occupied and dot
2 occupied with one electron with spin $\sigma$ (see appendix). A
remarkable result of Fig. 5 is the larger enhancement of the maximum
Seebeck coefficient. This enhancement of the maximum $S$ is due to
the degeneracy of QD levels, instead from larger $t_c$. Note that
$S$ is sensitive to the fluctuation of QD energy levels, but not to
$t_c$.$^{14)}$  This will be further demonstrated in the $N=5$ case.
Unlike two-QD junction with identical QDs, the spectra of these
thermal response functions for the three-QD junction do not possess
symmetric behavior as a result of the interdot Coulomb interactions.
The central dot (dot 2) feels the Coulomb interactions from both
dots 1 and 2, while dots 1 and 3 can only feel the interdot Coulomb
interaction from the central dot. Such an effect can be observed in
the behaviors of occupation numbers as functions of $eV_g$.

The electrical conductance $G_e$, Seebeck coefficient $S$, and
figure of merit ZT are plotted as functions of detuning energy
$\Delta=E_{\ell}-E_F$ with and without interdot Coulomb interaction
at $k_BT=10\Gamma_0$ in Fig.~6, where we have adopted
$t_{\ell,j}=6\Gamma_0$ and $\Gamma_l=\Gamma_R=\Gamma_0$. We note
$G_e$ is suppressed by the interdot Coulomb interactions. Such an
effect becomes weak, when $\Delta$ increases. $S$ is almost
independent of $U_{\ell,j}$ and it shows a linear dependence of
$\Delta$, roughly described by $-k_B\Delta/(eT)$. The behavior of ZT
as a function of $\Delta$ can be described by the function
ZT$=\alpha \Delta^2/(T^3 cosh^2(\Delta/2k_BT))$, where $\alpha$ is
independent of $T$ and $\Delta$. When $\Delta \gg k_BT$, $\kappa_e$
becomes negligible. Therefore, the thermal conductance is dominated
by $\kappa_{th}$, which is assumed to be a linear function of T. ZT
is determined by the power factor of $S^2G_e$. The maximum ZT occurs
near $\Delta_{max}=3k_BT$ and it is slightly reduced in the presence
of $U_{\ell,j}$. The effect of $U_{\ell,j}$ becomes negligible when
the QD energy levels are far above $E_F$. In the following
discussion, we will show that electron Coulomb interactions are
important when $E_F$ is above $E_{\ell}$.

Fig. 7 shows $G_e$, S, and ZT as functions of gate voltage at
$k_BT=10\Gamma_0$ and $E_{\ell}=E_0=E_F+50\Gamma_0-eV_g$ for various
interdot Coulomb interactions. Other physical parameters are the
same as those of Fig. 3.  Comparing with the low-temperature results
given in Fig. 5, we see that the six peaks of $G_e$ at
$k_BT=1\Gamma_0$ now become two broad peaks. We also note that the
presence of interdot Coulomb interactions breaks the structure
symmetry (taking $eV_g=80\Gamma_0$ as the mid pint, where $E_F$ sits
between $E_\ell$ and $E_\ell+U_\ell$ with equal separation). In
particular, $ZT_{max,>}$ (for QD energy levels above $E_F$) is only
slightly suppressed with increasing interdot Coulomb interactions.
On the other hand, $ZT_{max,<}$ (for QD energy levels below $E_F$)
is significantly suppressed. The difference in interdot Coulomb
interactions between central dot and two outer dots leads to an
artificial ``QD energy level fluctuation'' which suppresses ZT. Our
result indicates that the optimization of ZT prefers having the QD
levels above $E_F$, although many studies of individual QDs$^{11)}$
and two coupled QDs$^{12)}$ indicated that electron Coulomb
interactions can enhance ZT when QD levels are below $E_F$. Authors
in references [11] adopted the mean field approximation, which
truncates the high order Green's functions arising from electron
Coulomb interactions.

\subsection{Effect of number of QDs in SQDC}
Since in realistic QD junctions for thermoelectric application, one
needs to consider a large number of serially coupled QDs in order to
accommodate reasonable temperature gradient across the junction, it
is important to know the effect of number of QDs ($N$) on the
thermoelectric properties of a SQDC system for a given $t_c$.  To
clarify the effect of $N$ on ZT, we plot $G_e$, $S$, $\kappa_e$, and
ZT as functions of temperature for various values of $N$ by using
the following physical parameters: $E_{\ell}=E_0=E_F+30\Gamma_0$,
$U_{\ell}=U_0=30\Gamma_0$, $t_c=3\Gamma_0$. The interdot Coulomb
interactions have been turned off. Increasing the number of QDs
would change the density of states (DOS) of the SQDC system from
atomic limit to the band limit. For $N$ varying from 2 to 5, in the
absence of electron Coulomb interactions, we have the following
energy spectra
(under the condition $t_c/\Gamma \gg 1$)\\
\[ N=2: \epsilon=E_0+t_c,\; E_0-t_c; \]
\[ N=3: \epsilon=E_0+\sqrt{2}t_c, E_0, \; E_0-\sqrt{2}t_c; \]
\[ N=4: \epsilon=E_0+\sqrt{(3+\sqrt{5})/2}~t_c, \; E_0+\sqrt{(3-\sqrt{5})/2}~t_c, \]
 \[ E_0-\sqrt{(3-\sqrt{5})/2}~t_c, \; E_0-\sqrt{(3+\sqrt{5})/2}~t_c; \]
\[ N=5: \epsilon=E_0+\sqrt{3}t_c, \; E_0+t_c,\; E_0, \]
\[ E_0-t_c, \; E_0-\sqrt{3}t_c.\]
The separations between these resonant channels become small with
increasing QD numbers. In addition, these resonant channels have
different broadening. For example, three resonant channels of the
$N=3$ case occur at $\epsilon=E_0+\sqrt{2}t_c$, $E_0$, and
$E_0-\sqrt{2}t_c$, and their broadening widths are $\Gamma/4$,
$\Gamma/2$, and $\Gamma/4$, respectively. As expected, the maximum
ZT decreases with increasing $N$. Based on the results of Fig. 8,
the reduction of maximum ZT is attributed to the reduction of $G_e$.
Increasing $N$, the resonant channels increases whereas probability
weights of these resonant channels decrease. This explains why $G_e$
is reduced with increasing $N$. Note that when Coulomb interactions
are turned off, $G_e$ would become insensitive to $N$ for the case
of small $t_c$ and large $\Delta=E_{\ell}-E_F$. The larger $t_c$,
the more important the QD number effect becomes. In the high
temperature regime, we find that the Seebeck coefficient is not
sensitive to $N$  [see Fig. 8(b)], while $k_{e}$ reduces
significantly as $N$ increases [see Fig. 8(c)]. Thus, $k_{ph}$
dominates the thermal conductance in the large $N$ limit.
Consequently, the behavior of ZT is essentially determined by the
power factor ($pF=S^2G_e$), and the trend of ZT with respect to
increasing $N$ is similar to that of $G_e$. We also note that the
dependence of all thermoelectric functions on $N$ saturates once $N$
reaches 5 for the weak hopping strength considered, $t_c=3\Gamma_0$.
Thus, it is sufficient to model the thermoelectric behaviors of a
SQDC with large $N$ by using $N=5$. However, for larger $t_c$, the
saturation behavior would occur at larger $N$.

In order to find the optimization condition of ZT, we plot in Fig.~9
the occupation number of each QD, electrical conductance, Seebeck
coefficient, and ZT as functions of gate voltage for various
temperatures for a 5-dot SQDC. Other physical parameters are the
same as the $N=5$ case shown in Fig. 8. From Fig. 9(a) we see a
small difference in the occupation number between exterior dots
($N_1,N_5$) and interior dots ($N_2,N_3,N_4$) even though the
interdot Coulomb interactions are turned off. This phenomena also
occurs in the $N=3$ and $N=4$ cases (not shown here). The occupation
number fluctuation was also reported in Ref. [26] for the $N=3$ case
by solving the Master equation. For $eV_g=25\Gamma_0$, electrons
prefer to occupy the outer QDs ($N_1$ and $N_5$). For
$eV_g=75\Gamma_0$, electrons prefer to accumulate in the interior
QDs ($N_2,N_3,N_4$). The maximum electrical conductance is
suppressed with increasing temperature, while the peak width becomes
wider. Such a behavior is no different from a single QD with
multiple energy levels.$^{12)}$ However, we note that $G_{e,max}$
does not have a Lorentzian shape even though each resonant channel
has a Lorentzian shape. This is mainly attributed to the formation
of a "miniband" with the probability weight
$(1-N_{1,\bar\sigma})(1-N_{2\bar\sigma})(1-N_{3,\bar\sigma})(1-N_{4,\bar\sigma})(1-N_{5,\bar\sigma})$.
Note that once $U_{\ell,j}=0$, there are only two kind of
probability weights $(1-N_{\ell,\bar\sigma})$ and
$N_{\ell,\bar\sigma}$ for each QD.$^{30)}$ This characteristic can
also be demonstrated from the expressions for N=3 given in the
appendix. When the applied gate voltage continue to increase, the
SQDC turns into a Mott-insulator at half-filling
$N_{\ell}=N_{\ell,\sigma}=N_{\ell,\bar\sigma}=0.5$, resulting from
Coulomb band gap.$^{30)}$ (The electrical conductance almost
vanishes for each QD with one electron). The upper "miniband" with
probability weight
$N_{1,\bar\sigma}N_{2\bar\sigma}N_{3,\bar\sigma}N_{4,\bar\sigma}N_{5,\bar\sigma}$
arises for electrons hopping between energy levels at $E_{0}+U_{0}$.
The Seebeck coefficient shown in Fig.~9(c) goes through zero at
$V_{g1}$, $V_{g2}$, and $V_{g3}$, respectively. At these applied
gate voltages, the SQDC has an electron-hole symmetry. Note that the
Seebeck coefficient vanishes at the half-filling case. We see that
the electrical conductance and Seebeck coefficient are very
sensitive to QD energy levels. Due to the electron-hole symmetry,
the $G_e$ and ZT curves are symmetric (while $S$ curve is
antisymmetric) with respect to the mid point at $eV_g=45\Gamma_0$.
Thus, ZT curve has two maxima, one for QD energy levels above $E_F$
and the other for QD energy levels below $E_F$. The maximum ZT at
$k_BT=5\Gamma_0$ can reach 8 for $V_g$ near $10\Gamma_0$ and
$80\Gamma_0$. When $\kappa_{ph}$ dominates the thermal conductance,
the maximum ZT values are correlated to the maximum power factor.
The maximum ZT occurs at neither good conducting state nor
insulating state, because the maximum $G_e$ (good conductance)is
accompanied by a poor Seebeck coefficient, and vice versa. Note that
the results shown in Fig.~9 are for the case with no interdot
Coulomb interactions, i.e. $U_{\ell,j}=0$. Once $U_{\ell,j}$ are
turned on, the spectra of $G_e$ becomes somewhat complicated to
analyze. Meanwhile, it would significantly lower the maximum ZT
value for the peak with $E_F > E_0$.

Electron interdot hopping strength $t_c$ is a key parameter in
determining the bandwidth of the miniband, which would affect the
thermoelectric properties.$^{31)}$ Fig.~10 shows $G_e$, $S$,
$(ZT)_0$ and ZT as functions of $t_c$ for various detuning energies
$\Delta=E_{\ell}-E_F$ at $k_BT=10\Gamma_0$. Here, we consider the
case with on-site Coulomb interaction $U=60\Gamma_0$ and zero
interdot Coulomb interaction.  Note that $(ZT)_0$ corresponds to ZT
in the absence of phonon thermal conductance. When $t_c$ is smaller
than $\Gamma_0$, $G_e$ increases quickly with respect to $t_c$. This
behavior can be explained by the fact of that the transmission
factor is proportional to $t^8_c$ for the 5-dot case when $t_c$
approaches zero. Once $t_c$ is larger than $\Gamma_0$, $G_e$ becomes
almost saturated. $G_e$ values at $\Delta=10\Gamma_0$ are very close
to those at $\Delta=20\Gamma_0$. Such a behavior is also seen in the
dotted line of Fig.~9(b), in which the QD energy level is tuned by
the gate voltage. We notice that $S$ is rather insensitive to $t_c$
in the weak hoping limit, $t_c/\Gamma_0 \ll 1$ ($S \approx
-k_B\Delta/eT$). In the absence of phonon thermal conductance,
$(ZT)_0$ diverges in the weak hoping limit, because the Lorenz
number $L=\kappa_e/(G_e T)$ approaches zero. From Fig.~10(d), we see
that ZT increases with increasing $t_C$, reaching an optimum value
at $t_C \approx 2\Gamma_0$ and then decreases for higher $t_c$. The
reduction of ZT at higher $t_c$ arises from the faster reduction of
$S^2$ in comparison with the increase of $G_e$. To reveal how the
asymmetrical coupling between the QDs and the electrodes influences
ZT, we also plot ZT versus $V_g$ for the detuning energy
$\Delta=30\Gamma_0$ for different tunneling rates in Fig.~10(d).
(See curves marked by filled triangles and diamonds) The maximum ZT
is suppressed when the ratio of $\Gamma_L/\Gamma_R$ is far away from
1, while keeping the same average value,
$\Gamma_L+\Gamma_R=2\Gamma_0$.

\section{Summary and conclusions}
We have theoretically investigated the thermoelectric effects of
SQDC connected to metallic electrodes. The length of the SQDC is
finite, and shorter than the electron mean free path. Thus, the
electron-phonon interaction effect can be ignored in this study.
{\color{blue} Once the length of nanowire is long enough, the inelastic
scattering of electrons becomes important, which will seriously suppress the optimal ZT values.$^{12)}$}
We have derived
an expression for the transmission coefficient ${\cal T}(\epsilon)$
in terms of the on-site retarded Green's functions and effective
tunneling rates involving electron hopping and Coulomb interactions,
which keeps the same form as the Landauer formula for charge and
heat currents.

In the linear response regime, electrical conductance $G_e$, Seebeck
coefficient $S$, electron thermal conductance $\kappa_e$, and figure
of merit ZT are calculated for finite length SQDCs with QD number
ranging from 2 to 5. The thermoelectric properties of spin-dependent
configurations (spin triplet and singlet states) are studied. The
nonthermal broadening behavior of $G_e$ is maintained in the
three-QD case in the presence of electron Coulomb interactions,
because such a behavior is the essential feature of resonant
tunneling junction system. Unlike the case of double QDs,$^{14)}$
the interdot Coulomb interactions lead to considerable suppression
of the maximum ZT value in the three-QD case when the degenerate QD
energy levels are below $E_F$. This is caused by the symmetry
breaking via interdot Coulomb interactions in the SQDC junction.

We find that for a given $t_c$, increasing the QD number $N$ in the
SQDC would suppress the maximum ZT value, and the reduction quickly
saturates once $N$ reaches 5 in the weak hopping limit ($t_c/U \ll
1$). However, for larger $t_c$, the saturation behavior would occur
at larger $N$. In addition, we find that the Seebeck coefficient is
insensitive to $t_c$ and $U_{\ell}$($U_{\ell,j}$) when QD energy
levels are far above $E_F$, and it can be approximated by a simple
linear expression, $S=-k_B\Delta/eT$, where $\Delta=E_{\ell}-E_F$
and T is the equilibrium temperature. This characteristic of
$S=-k_B\Delta/eT$ was also reported in an infinite Hubbard chain
with narrow bandwidth.$^{32)}$ The feature of $S=\Delta V/\Delta
T\approx -k_B\Delta/eT$ can be utilized to realize a temperature
sensor.$^{33)}$ High ZT values are possibly achieved by tailoring
the intradot and interdot Coulomb interactions as well as by
detuning the energy difference between the energy levels of the QDs
and the Fermi energy level of the electrodes.

\section*{Acknowledgment}
This work was supported in part by  National Science Council, Taiwan
under Contract Nos. NSC 101-2112-M-008-014-MY2 and NSC
101-2112-M-001-024-MY3.
\mbox{}\\
$^{\dagger}$ E-mail address: mtkuo@ee.ncu.edu.tw\\
$^*$ E-mail address: yiachang@gate.sinica.edu.tw

\appendix

\section{transmission coefficient of N=3}

{\color{blue}In this appendix, we give the detailed expression of
${\cal T}_{1,3}$ for the N=3 case. For simplicity, we assume the
same interdot hopping constant between any two QDs, i.e.
$t_{\ell,j}=t_c$, and we denote $U_{1,2}=U_{2,1}\equiv U_I$,
$U_{2,3}=U_{3,2}\equiv U_J$, and
$\mu_{\ell}=\epsilon-E_{\ell}+i\Gamma_{\ell}/2$; $\ell=1,2,3$. Note
that $\Gamma_2=0$, because QD 2 is not directly coupled to the
electrodes. Eq. (7) becomes
\begin{equation}
{\cal T}_{1,3}(\epsilon)=-2\sum^{32}_{m=1}\frac{\Gamma_{1}(\epsilon)
\Gamma^{m}_{1,3}(\epsilon)}{\Gamma_{1}(\epsilon)+\Gamma^{m}_{1,3}(\epsilon)}
\mbox{Im}G^r_{1,m}(\epsilon),
\end{equation}
where
\begin{equation}
G^r_{1,m}(\epsilon)=\frac{p_{1,m}}{\mu_{1}-\Pi_{1,m}-\Sigma^m_{1,3}},
\end{equation}
\begin{equation}
\Sigma^m_{1,3}=\frac{t^2_c}{\mu_2-\Pi_{2,m}-\frac{t^2_c}{\mu_3-\Pi_{3,m}}},
\end{equation}
and $\Gamma^m_{1,3}=-2 Im\Sigma^m_{1,3}$ denotes the effective
tunneling rate of the electron from the first QD energy level to the
right electrode via channel $m$. The probability weights ($p_{1,m}$)
and Coulomb energies ($\Pi_{\ell,m}$) are given by

\[ p_{1,1} = \bar a_1a_2a_3; \Pi_{1,1}=\Pi_{2,1}=\Pi_{3,1}=0. \]
\[ p_{1,2} = \bar a_1a_2 b_{3,\bar\sigma}; \Pi_{1,2}=0, \Pi_{2,2}=U_{J}, \Pi_{3,2}=U_3. \]
\[ p_{1,3} = \bar a_1a_2 b_{3,\sigma}; \Pi_{1,3}=0, \Pi_{2,3}=\Pi_{3,3}=U_{J}. \]
\[ p_{1,4} = \bar a_1a_2c_{3}; \Pi_{1,4}=0, \Pi_{2,4}=2U_{J}, \Pi_{3,4}=U_3+U_{J}. \]
\[ p_{1,5} = \bar a_1b_{2,\bar\sigma} a_3; \Pi_{1,5}=U_I, \Pi_{2,5}=U_{2}, \Pi_{3,5}=U_{J}. \]
\[ p_{1,6} = \bar a_1b_{2,\bar\sigma} b_{3,\bar\sigma}; \Pi_{1,6}=U_I, \Pi_{2,6}=U_{2}+U_J, \Pi_{3,6}=U_3+U_{J}. \]
\[ p_{1,7} = \bar a_1b_{2,\bar\sigma}b_{3,\sigma}; \Pi_{1,7}=U_I, \Pi_{2,7}=U_{2}+U_J, \Pi_{3,7}=2U_{J}. \]
\[ p_{1,8} = \bar a_1b_{2,\bar\sigma}c_{3};\Pi_{1,8}=U_I, \Pi_{2,8}=U_{2}+2U_J, \Pi_{3,8}=U_3+2U_{J}. \]
\[ p_{1,9} = \bar a_1b_{2,\sigma}a_3; \Pi_{1,9}=U_I, \Pi_{2,9}=U_I, \Pi_{3,9}=0. \]
\[ p_{1,10} = \bar a_1b_{2,\sigma}b_{3,\bar\sigma}; \Pi_{1,10}=U_I, \Pi_{2,10}=U_I+U_J, \Pi_{3,10}=U_3. \]
\[ p_{1,11} = \bar a_1b_{2,\sigma}b_{3,\sigma}; \Pi_{1,11}=U_I, \Pi_{2,11}=U_I+U_J, \Pi_{3,11}=U_J. \]
\[ p_{1,12} = \bar a_1b_{2,\sigma}c_{3}; \Pi_{1,12}=U_I, \Pi_{2,12}=U_I+2U_J, \Pi_{3,12}=U_3+U_J. \]
\[ p_{1,13} = \bar a_1c_{2}a_3;  \Pi_{1,13}=2U_I, \Pi_{2,13}=U_2+U_I, \Pi_{3,13}=U_J. \]
\[ p_{1,14} = \bar a_1c_{2} b_{3,\bar\sigma}; \]\[ \Pi_{1,14}=2U_I, \Pi_{2,14}=U_2+U_I+U_J, \Pi_{3,14}=U_3+U_J. \]
\[ p_{1,15} = \bar a_1c_{2}b_{3,\sigma}; \]\[ \Pi_{1,15}=2U_I, \Pi_{2,15}=U_2+U_I+U_J, \Pi_{3,15}=2U_J. \]
\[ p_{1,16} = \bar a_1c_{2}c_{3}; \]\[ \Pi_{1,16}=2U_I, \Pi_{2,16}=U_2+U_I+2U_J, \Pi_{3,16}=U_3+2U_J. \]

\[ p_{1,17} = N_{1,\bar\sigma}a_2a_3; \Pi_{1,17}=U_1,\Pi_{2,17}=U_I,\Pi_{3,17}=0. \]
\[ p_{1,18} = N_{1,\bar\sigma}a_2 b_{3,\bar\sigma}; \Pi_{1,18}=U_1, \Pi_{2,18}=U_I+U_{J}, \Pi_{3,18}=U_3. \]
\[ p_{1,19} = N_{1,\bar\sigma}a_2 b_{3,\sigma}; \Pi_{1,19}=U_1, \Pi_{2,19}=U_I+U_J,\Pi_{3,19}=U_{J}. \]
\[ p_{1,20} = N_{1,\bar\sigma}a_2c_{3}; \]\[\Pi_{1,20}=U_1, \Pi_{2,20}=U_I+2U_{J}, \Pi_{3,20}=U_3+U_{J}. \]
\[ p_{1,21} = N_{1,\bar\sigma}b_{2,\bar\sigma} a_3; \]\[\Pi_{1,21}=U_1+U_I, \Pi_{2,21}=U_{2}+U_I, \Pi_{3,21}=U_{J}. \]
\[ p_{1,22} = N_{1,\bar\sigma}b_{2,\bar\sigma} b_{3,\bar\sigma}; \]\[\Pi_{1,22}=U_1+U_I, \Pi_{2,22}=U_{2}+U_I+U_J, \Pi_{3,22}=U_3+U_{J}. \]
\[ p_{1,23} = N_{1,\bar\sigma}b_{2,\bar\sigma}b_{3,\sigma}; \]\[\Pi_{1,23}=U_1+U_I, \Pi_{2,23}=U_{2}+U_I+U_J, \Pi_{3,23}=2U_{J}. \]
\[ p_{1,24} = N_{1,\bar\sigma}b_{2,\bar\sigma}c_{3};\] \[ \Pi_{1,24}=U_1+U_I, \Pi_{2,24}=U_{2}+U_I+2U_J, \Pi_{3,24}=U_3+2U_{J}. \]
\[ p_{1,25} = N_{1,\bar\sigma}b_{2,\sigma}a_3; \] \[  \Pi_{1,25}=U_1+U_I, \Pi_{2,25}=2U_I, \Pi_{3,25}=0. \]
\[ p_{1,26} = N_{1,\bar\sigma}b_{2,\sigma}b_{3,\bar\sigma}; \] \[\Pi_{1,26}=U_1+U_I, \Pi_{2,26}=2U_I+U_J, \Pi_{3,26}=U_3. \]
\[ p_{1,27} = N_{1,\bar\sigma}b_{2,\sigma}b_{3,\sigma}; \]\[ \Pi_{1,27}=U_1+U_I, \Pi_{2,27}=2U_I+U_J, \Pi_{3,27}=U_J. \]
\[ p_{1,28} = N_{1,\bar\sigma}b_{2,\sigma}c_{3}; \]\[\Pi_{1,28}=U_1+U_I, \Pi_{2,28}=2U_I+2U_J, \Pi_{3,28}=U_3+U_J. \]
\[ p_{1,29} = N_{1,\bar\sigma}c_{2}a_3; \]\[ \Pi_{1,29}=U_1+2U_I, \Pi_{2,29}=U_2+2U_I, \Pi_{3,29}=U_J. \]
\[ p_{1,30} = N_{1,\bar\sigma}c_{2} b_{3,\bar\sigma}; \]\[ \Pi_{1,30}=U_1+2U_I, \Pi_{2,30}=U_2+2U_I+U_J, \Pi_{3,30}=U_3+U_J. \]
\[ p_{1,31} = N_{1,\bar\sigma}c_{2}b_{3,\sigma}; \]\[ \Pi_{1,31}=U_1+2U_I, \Pi_{2,31}=U_2+2U_I+U_J, \Pi_{3,31}=2U_J. \]
\[ p_{1,32} = N_{1,\bar\sigma}c_{2}c_{3}; \]\[ \Pi_{1,32}=U_1+2U_I, \Pi_{2,32}=U_2+2U_I+2U_J, \Pi_{3,32}=U_3+2U_J. \]
For the retarded Green's unction of QD 2, we have
\begin{eqnarray}
G^r_{2,m}(\epsilon)&=&\frac{p_{2,m}}{\mu_2-\Pi_{2,m}-\Sigma^{m}_{2,1}-\Sigma^{m}_{2,3}
},
\end{eqnarray}
where  $\Sigma^m_{2,1}=\frac{t^2_c}{\mu_1-\Pi_{1,m}}$ and
$\Sigma^m_{2,3}=\frac{t^2_c}{\mu_3-\Pi_{3,m}}$. The probability
factors, $p_{2,m}$ and Coulomb energies $\Pi_{\ell,m}; \ell=1,2,3$
are given by

\[ p_{2,1} = \bar a_2a_1a_3; \Pi_{1,1}=\Pi_{2,1}=\Pi_{3,1}=0. \]
\[ p_{2,2} = \bar a_2a_1 b_{3,\bar\sigma}; \Pi_{1,2}=0, \Pi_{2,2}=U_{J}, \Pi_{3,2}=U_3. \]
\[ p_{2,3} = \bar a_2a_1 b_{3,\sigma}; \Pi_{1,3}=0, \Pi_{2,3}=\Pi_{3,3}=U_{J}. \]
\[ p_{2,4} = \bar a_2a_1c_{3}; \Pi_{1,4}=0, \Pi_{2,4}=2U_{J}, \Pi_{3,4}=U_3+U_{J}. \]
\[ p_{2,5} = \bar a_2b_{1,\bar\sigma} a_3; \Pi_{1,5}=U_1, \Pi_{2,5}=U_{I}, \Pi_{3,5}=0. \]
\[ p_{2,6} = \bar a_2b_{1,\bar\sigma} b_{3,\bar\sigma}; \Pi_{1,6}=U_1, \Pi_{2,6}=U_{I}+U_J, \Pi_{3,6}=U_3. \]
\[ p_{2,7} = \bar a_2b_{1,\bar\sigma}b_{3,\sigma}; \Pi_{1,7}=U_1, \Pi_{2,7}=U_{I}+U_J, \Pi_{3,7}=U_{J}. \]
\[ p_{2,8} = \bar a_2b_{1,\bar\sigma}c_{3};\Pi_{1,8}=U_1, \Pi_{2,8}=U_{I}+2U_J, \Pi_{3,8}=U_3+U_{J}. \]
\[ p_{2,9} = \bar a_2b_{1,\sigma}a_3; \Pi_{1,9}=U_I, \Pi_{2,9}=U_I, \Pi_{3,9}=0. \]
\[ p_{2,10} = \bar a_2b_{1,\sigma}b_{3,\bar\sigma}; \Pi_{1,10}=U_I, \Pi_{2,10}=U_I+U_J, \Pi_{3,10}=U_3. \]
\[ p_{2,11} = \bar a_2b_{1,\sigma}b_{3,\sigma}; \Pi_{1,11}=U_I, \Pi_{2,11}=U_I+U_J, \Pi_{3,11}=U_J. \]
\[ p_{2,12} = \bar a_2b_{1,\sigma}c_{3}; \Pi_{1,12}=U_I, \Pi_{2,12}=U_I+2U_J, \Pi_{3,12}=U_3+U_J. \]
\[ p_{2,13} = \bar a_2c_{1}a_3;  \Pi_{1,13}=U_1+U_I, \Pi_{2,13}=2U_I, \Pi_{3,13}=0. \]
\[ p_{2,14} = \bar a_2c_{1} b_{3,\bar\sigma}; \]\[ \Pi_{1,14}=U_1+U_I, \Pi_{2,14}=2U_I+U_J, \Pi_{3,14}=U_3. \]
\[ p_{2,15} = \bar a_2c_{1}b_{3,\sigma}; \]\[ \Pi_{1,15}=U_1+U_I, \Pi_{2,15}=2U_I+U_J, \Pi_{3,15}=U_J. \]
\[ p_{2,16} = \bar a_2c_{1}c_{3}; \]\[ \Pi_{1,16}=U_1+U_I, \Pi_{2,16}=2U_I+2U_J, \Pi_{3,16}=U_3+U_J. \]
\[ p_{2,17} = N_{2,\bar\sigma}a_1a_3; \Pi_{1,17}=U_I,\Pi_{2,17}=U_2,\Pi_{3,17}=U_J. \]
\[ p_{2,18} = N_{2,\bar\sigma}a_1 b_{3,\bar\sigma}; \Pi_{1,18}=U_I, \Pi_{2,18}=U_2+U_{J}, \Pi_{3,18}=U_3+U_J. \]
\[ p_{2,19} = N_{2,\bar\sigma}a_1 b_{3,\sigma}; \Pi_{1,19}=U_I, \Pi_{2,19}=U_2+U_J,\Pi_{3,19}=2U_{J}. \]
\[ p_{2,20} = N_{2,\bar\sigma}a_1c_{3}; \]\[\Pi_{1,20}=U_I, \Pi_{2,20}=U_2+2U_{J}, \Pi_{3,20}=U_3+2U_{J}. \]
\[ p_{2,21} = N_{2,\bar\sigma}b_{1,\bar\sigma} a_3; \]\[\Pi_{1,21}=U_1+U_I, \Pi_{2,21}=U_{2}+U_I, \Pi_{3,21}=U_{J}. \]
\[ p_{2,22} = N_{2,\bar\sigma}b_{1,\bar\sigma} b_{3,\bar\sigma}; \]\[\Pi_{1,22}=U_1+U_I, \Pi_{2,22}=U_{2}+U_I+U_J, \Pi_{3,22}=U_3+U_{J}. \]
\[ p_{2,23} = N_{2,\bar\sigma}b_{1,\bar\sigma}b_{3,\sigma}; \]\[\Pi_{1,23}=U_1+U_I, \Pi_{2,23}=U_{2}+U_I+U_J, \Pi_{3,23}=2U_{J}. \]
\[ p_{2,24} = N_{2,\bar\sigma}b_{1,\bar\sigma}c_{3};\] \[ \Pi_{1,24}=U_1+U_I, \Pi_{2,24}=U_{2}+U_I+2U_J, \Pi_{3,24}=U_3+2U_{J}. \]
\[ p_{2,25} = N_{2,\bar\sigma}b_{1,\sigma}a_3; \] \[  \Pi_{1,25}=2U_I, \Pi_{2,25}=U_2+U_I, \Pi_{3,25}=U_J. \]
\[ p_{2,26} = N_{2,\bar\sigma}b_{1,\sigma}b_{3,\bar\sigma}; \] \[\Pi_{1,26}=2U_I, \Pi_{2,26}=U_2+U_I+U_J, \Pi_{3,26}=U_3+U_J. \]
\[ p_{2,27} = N_{2,\bar\sigma}b_{1,\sigma}b_{3,\sigma}; \]\[ \Pi_{1,27}=2U_I, \Pi_{2,27}=U_2+U_I+U_J, \Pi_{3,27}=2U_J. \]
\[ p_{2,28} = N_{2,\bar\sigma}b_{1,\sigma}c_{3}; \]\[\Pi_{1,28}=2U_I, \Pi_{2,28}=U_2+U_I+2U_J, \Pi_{3,28}=U_3+2U_J. \]
\[ p_{2,29} = N_{2,\bar\sigma}c_{1}a_3; \]\[ \Pi_{1,29}=U_1+2U_I, \Pi_{2,29}=U_2+2U_I, \Pi_{3,29}=U_J. \]
\[ p_{2,30} = N_{2,\bar\sigma}c_{1} b_{3,\bar\sigma}; \]\[ \Pi_{1,30}=U_1+2U_I, \Pi_{2,30}=U_2+2U_I+U_J, \Pi_{3,30}=U_3+U_J. \]
\[ p_{2,31} = N_{2,\bar\sigma}c_{1}b_{3,\sigma}; \]\[ \Pi_{1,31}=U_1+2U_I, \Pi_{2,31}=U_2+2U_I+U_J, \Pi_{3,31}=2U_J. \]
\[ p_{2,32} = N_{2,\bar\sigma}c_{1}c_{3}; \]\[ \Pi_{1,32}=U_1+2U_I, \Pi_{2,32}=U_2+2U_I+2U_J, \Pi_{3,32}=U_3+2U_J. \]
$G^r_{3,m}(\epsilon)$ is obtained from $G^r_{1,m}(\epsilon)$ by
exchanging the indices 1 and 3. The average occupation numbers
$N_{\ell,\sigma}$, $N_{\ell,\bar\sigma}$, and  $c_{\ell}$ are
determined by solving Eqs. (18) and (19). Note that in the
uncorrelated limit ($U_{\ell}$ = 0 and $U_{ij}$ = 0), the expression
of Eq. (A2) can be obtained by the recursion method.$^{34)}$}


\newpage

\section*{Figures and Figure captions}

\begin{figure}[h]
\includegraphics[scale=0.4]{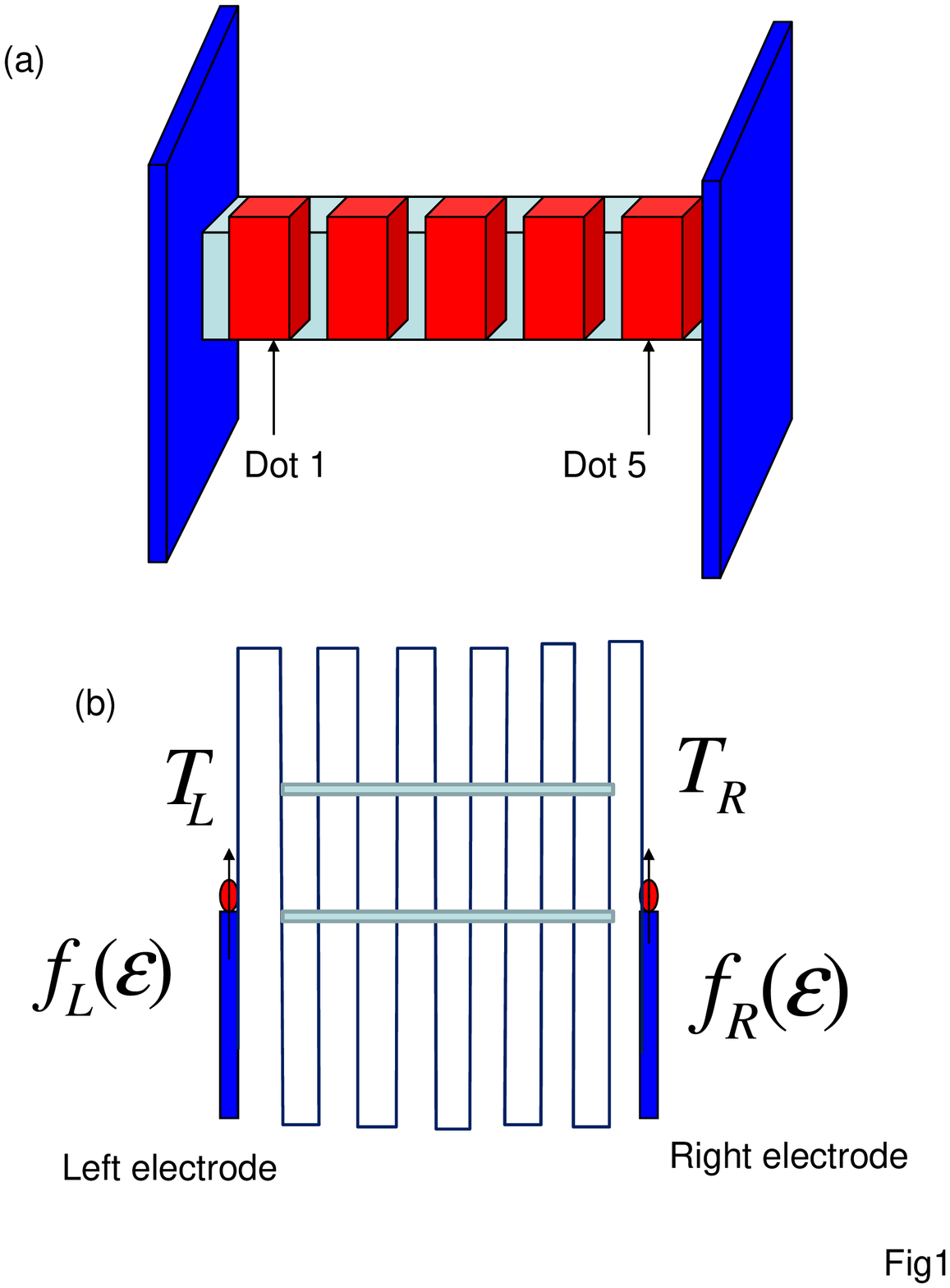}
\centering \caption{(a) Schematics for a semiconductor quantum dot
molecule (SQDC) connected to metallic electrodes, where $T_L$ and
$T_R$ describe the equilibrium temperature of the left and right
electrodes, respectively. (b) The band diagram illustrating a SQDC
connected to metallic electrodes.}
\end{figure}

\begin{figure}[h]
\includegraphics[scale=0.4]{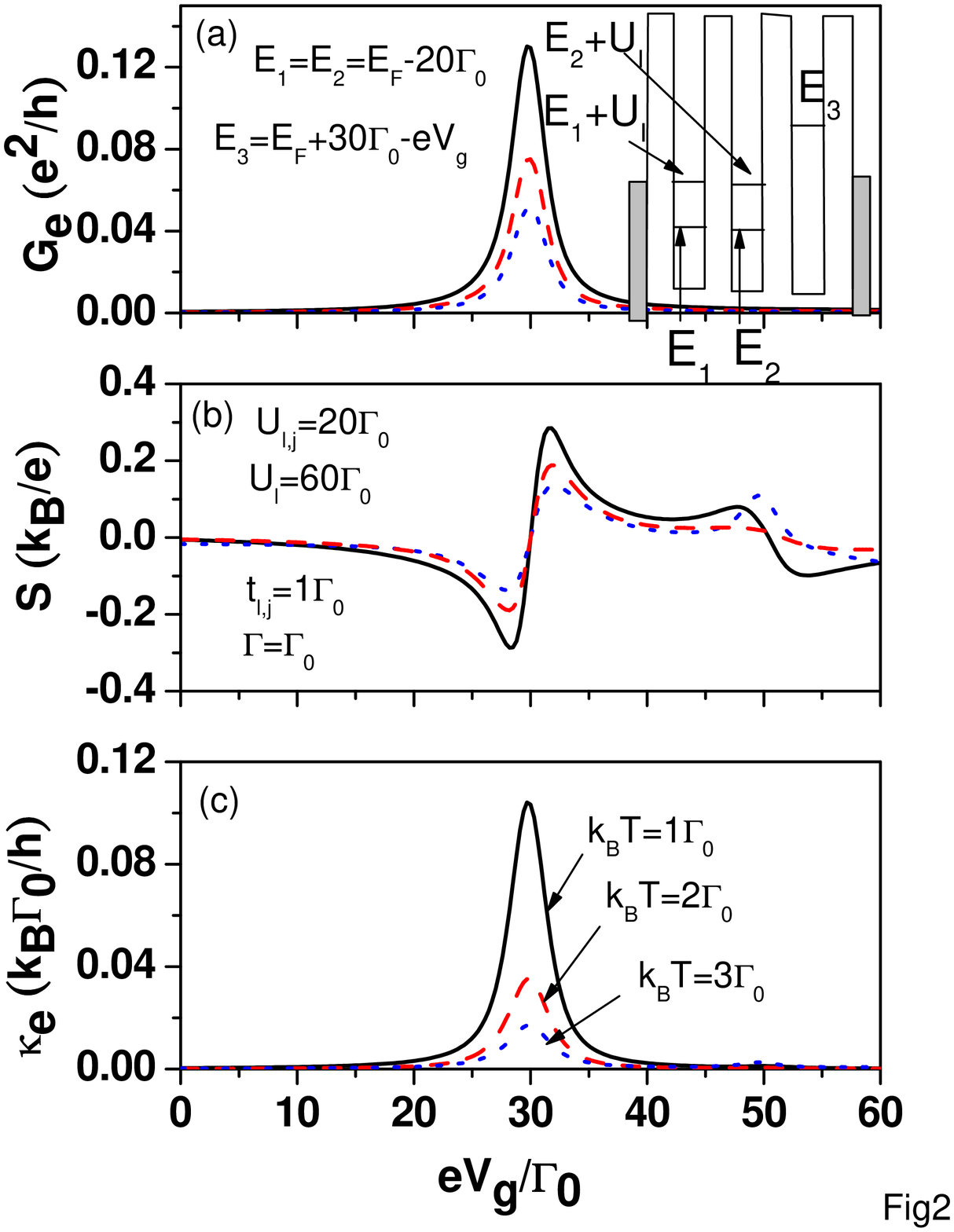}
\centering \caption{Electrical conductance ($G_e$), Seebeck
coefficient (S) and electrical thermal conductance ($\kappa_e$) as
functions of gate voltage ($V_g$), which is used to tune the $E_3$
level  for various temperatures. $E_1=E_2=E_F-20\Gamma_0$,
$E_3=E_F+30\Gamma_0-eV_g$. $U_{\ell}=60\Gamma_0$,
$U_{\ell,j}=20\Gamma_0$, $t_{\ell,j}=1\Gamma_0$, and
$\Gamma_L=\Gamma_R=\Gamma_0$}
\end{figure}

\begin{figure}[h]
\centering
\includegraphics[scale=0.4]{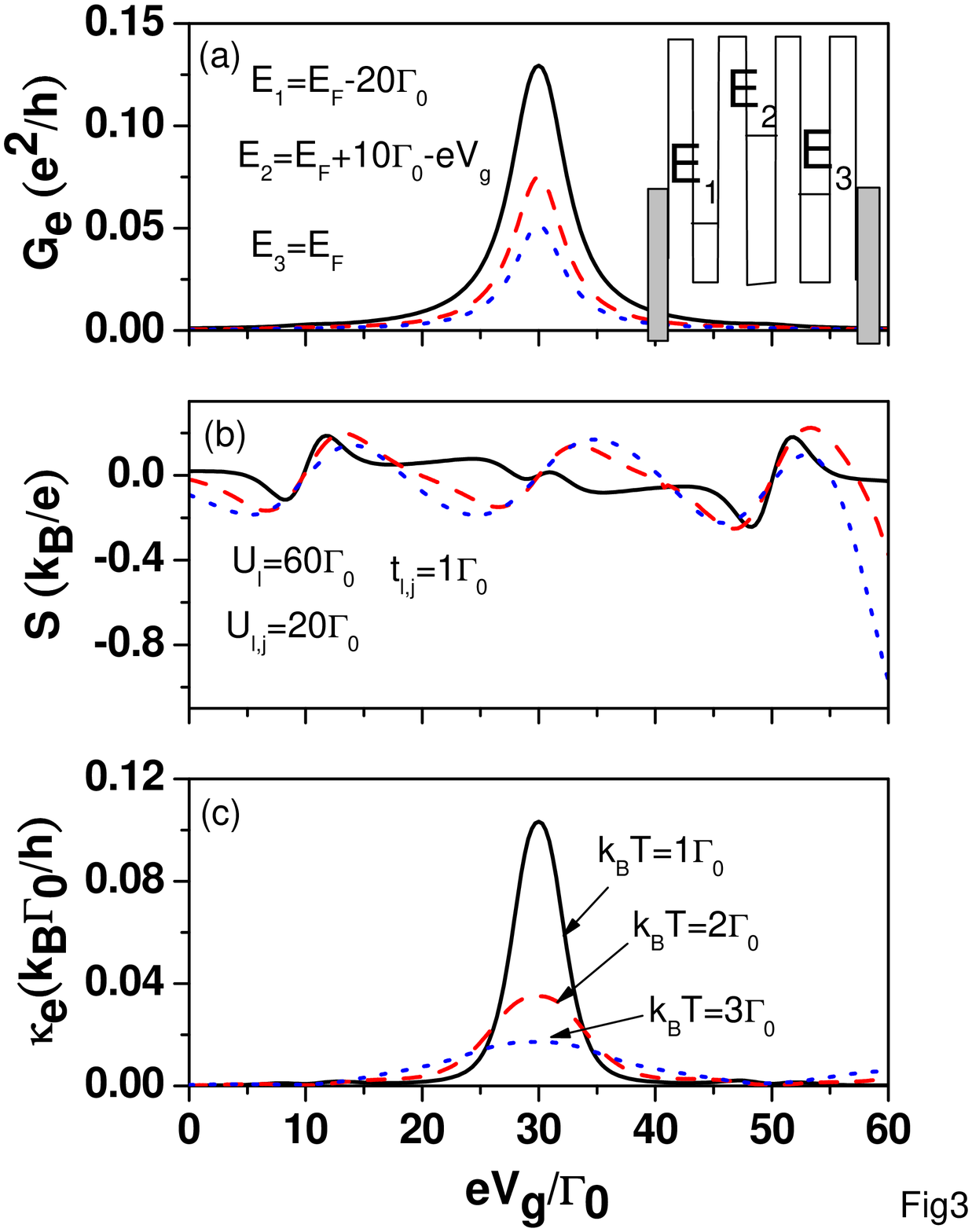}
\caption{Electrical conductance ($G_e$), Seebeck coefficient ($S$)
and electrical thermal conductance ($\kappa_e$) as functions of gate
voltage ($V_g$), which is used to tune the $E_2$ level for various
temperatures. $E_1=E_F-20\Gamma_0$, $E_2=E_F+10\Gamma_0-eV_g$, and
$E_3=E_F$. Other physical parameters are the same as those of
Fig.~2.}
\end{figure}

\begin{figure}[h]
\centering
\includegraphics[scale=0.4]{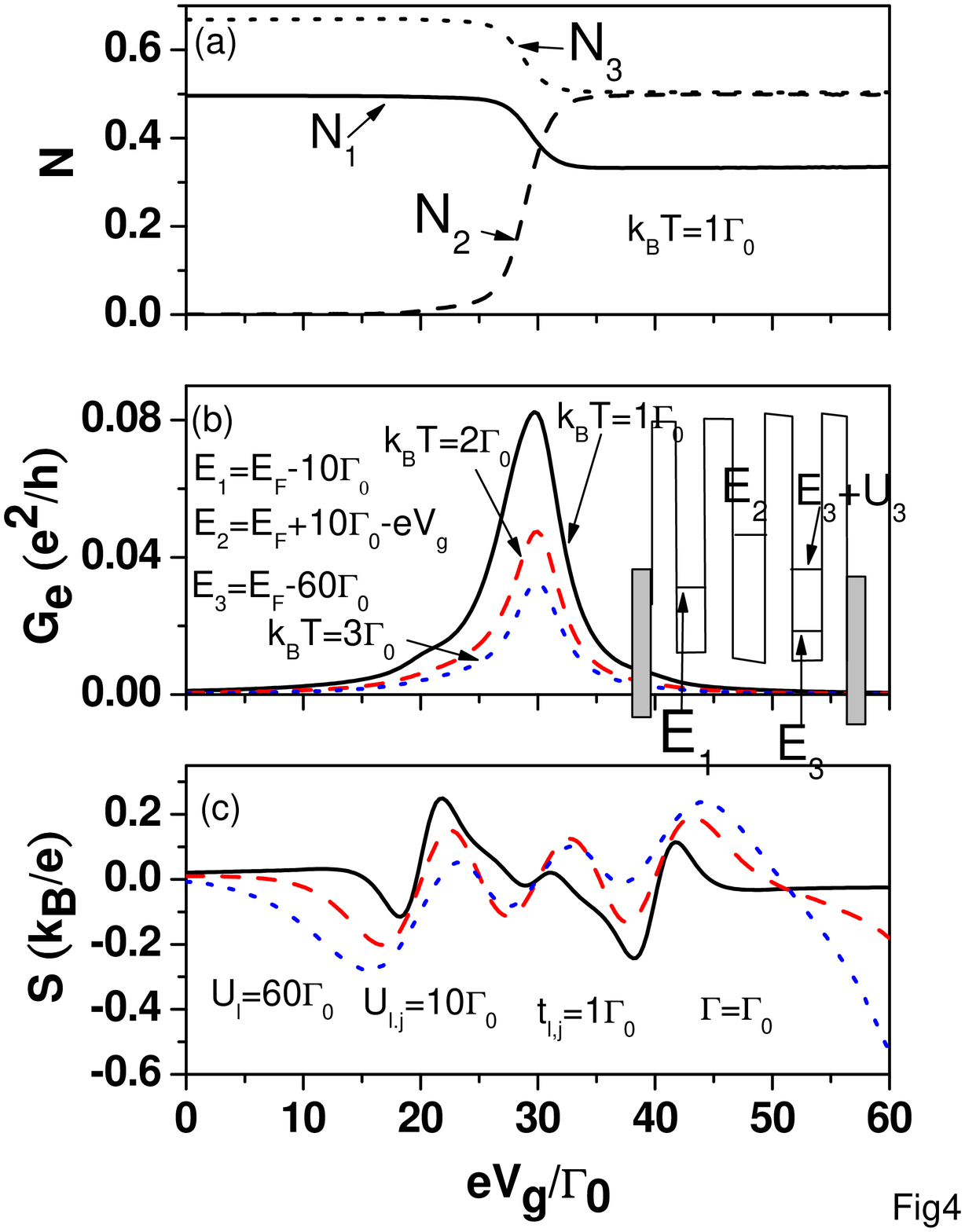}
\caption{ Occupation number ($N$), electrical conductance ($G_e$),
and Seebeck coefficient ($S$) as functions of gate voltage ($V_g$),
which is used to tune the $E_2$ level for various temperatures.
$E_1=E_F-10\Gamma_0$, $E_2=E_F+10\Gamma_0-eV_g$, and
$E_3=E_F-60\Gamma_0$. $U_{\ell,j}=10\Gamma_0$. Other physical
parameters are the same as those of Fig.~2}
\end{figure}

\begin{figure}[h]
\centering
\includegraphics[scale=0.3]{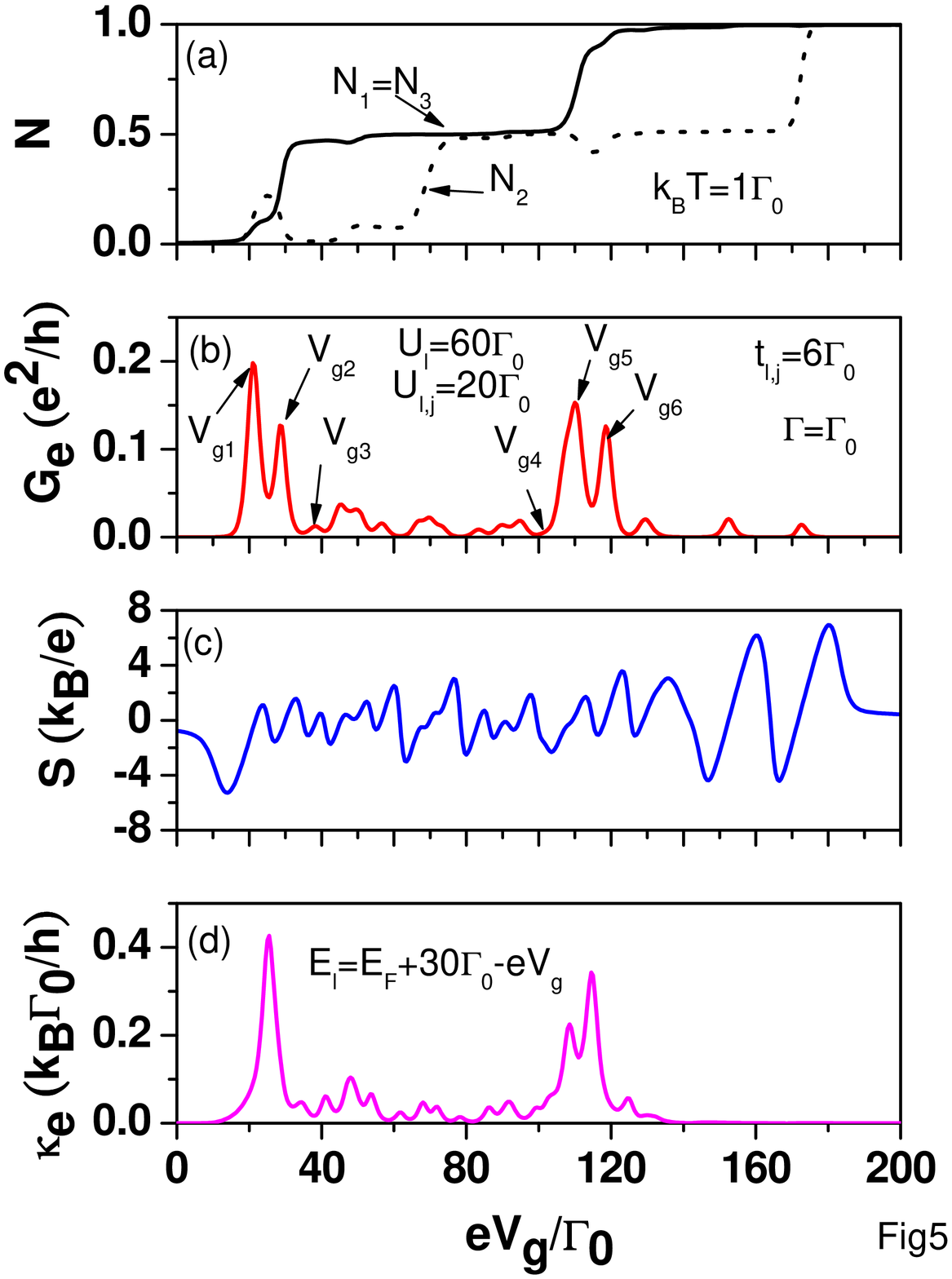}
\caption{Occupation number ($N$), electrical conductance ($G_e$),
Seebeck coefficient ($S$), and electrical thermal conductance
($\kappa_e$) as functions of gate voltage ($V_g$) at low temperature
($k_BT=1\Gamma_0$) for  QDs with identical energy levels
$E_{\ell}=E_F+30\Gamma_0-eV_g$ and $t_{\ell,j}=6\Gamma_0$. Other
physical parameters are the same as those of Fig. 2. }
\end{figure}

\begin{figure}[h]
\centering
\includegraphics[scale=0.4]{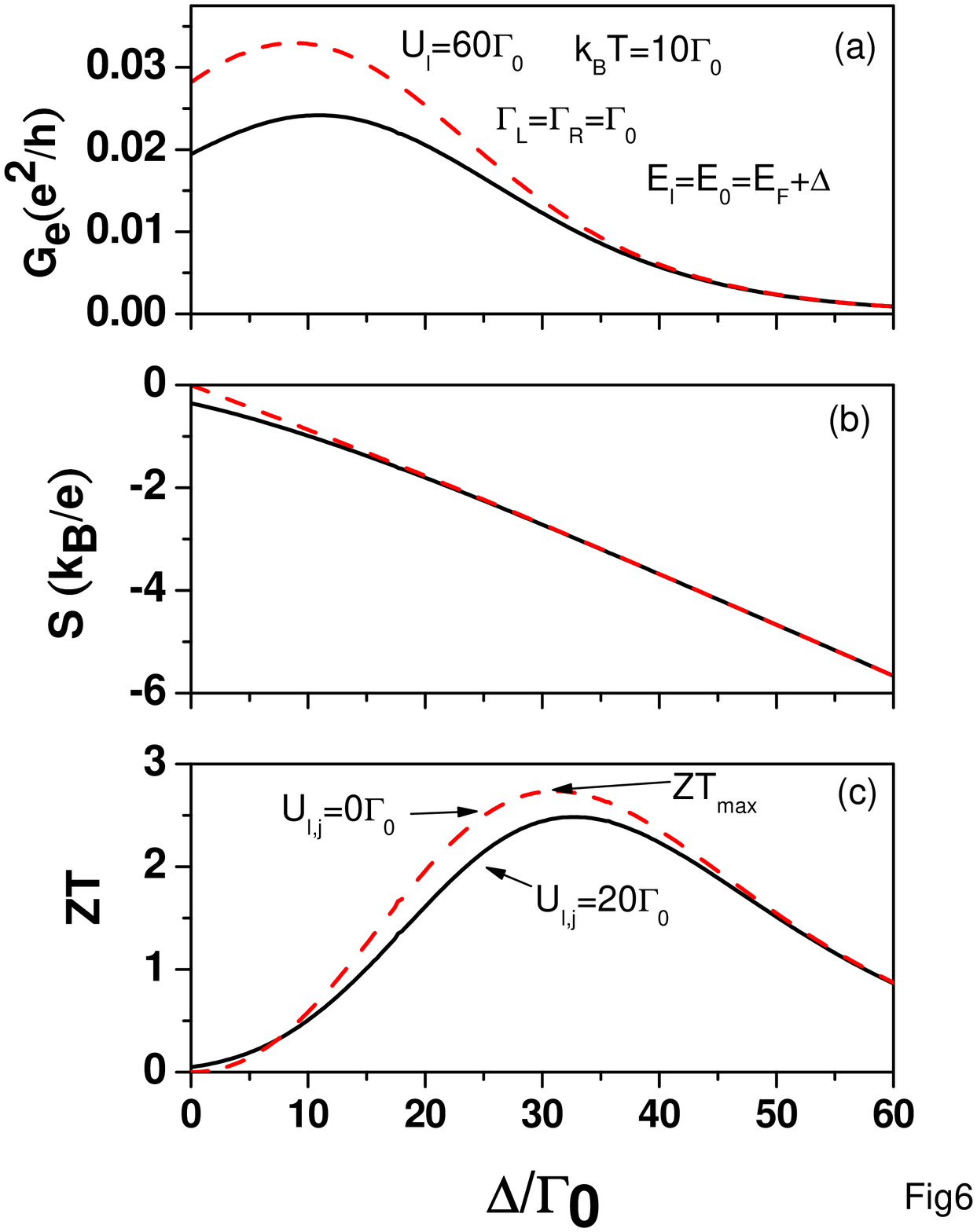}
\caption{Electrical conductance ($G_e$), Seebeck coefficient ($S$),
and ZT as functions of detuning energy, $\Delta=E_{\ell}-E_F$ at
temperature $k_BT=10\Gamma_0$ with and without interdot Coulomb
interactions.}
\end{figure}

\begin{figure}[h]
\centering
\includegraphics[scale=0.4]{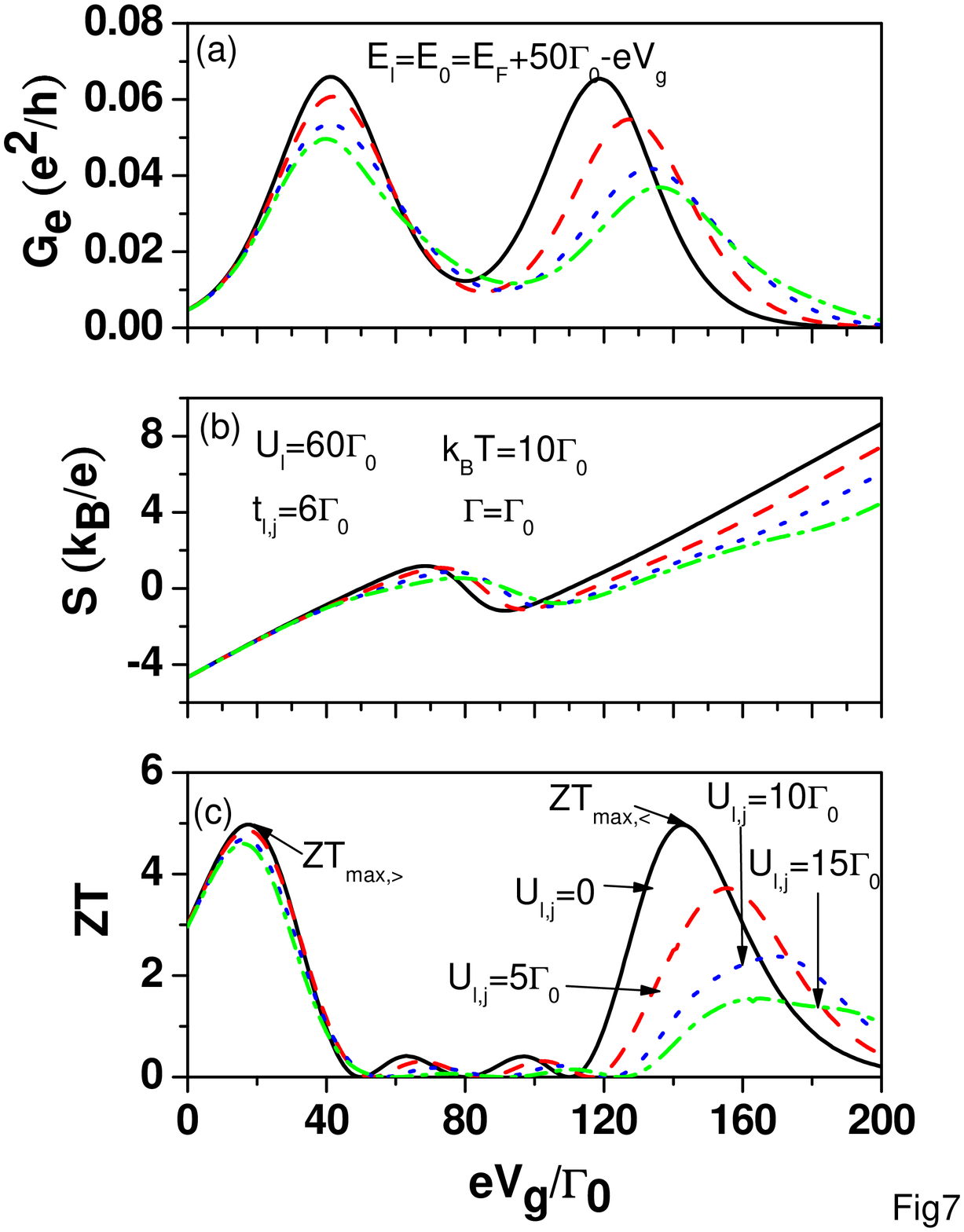}
\caption{Electrical conductance $G_e$, Seebeck coefficient (S), and
ZT as functions of gate voltage ($V_g$) at temperature
$k_BT=10\Gamma_0$ for various interdot Coulomb interactions.
$E_{\ell}=E_F+50\Gamma_0-eV_g$. Other physical parameters are the
same as those of Fig. 5.
 }
\end{figure}

\begin{figure}[h]
\centering
\includegraphics[scale=0.4]{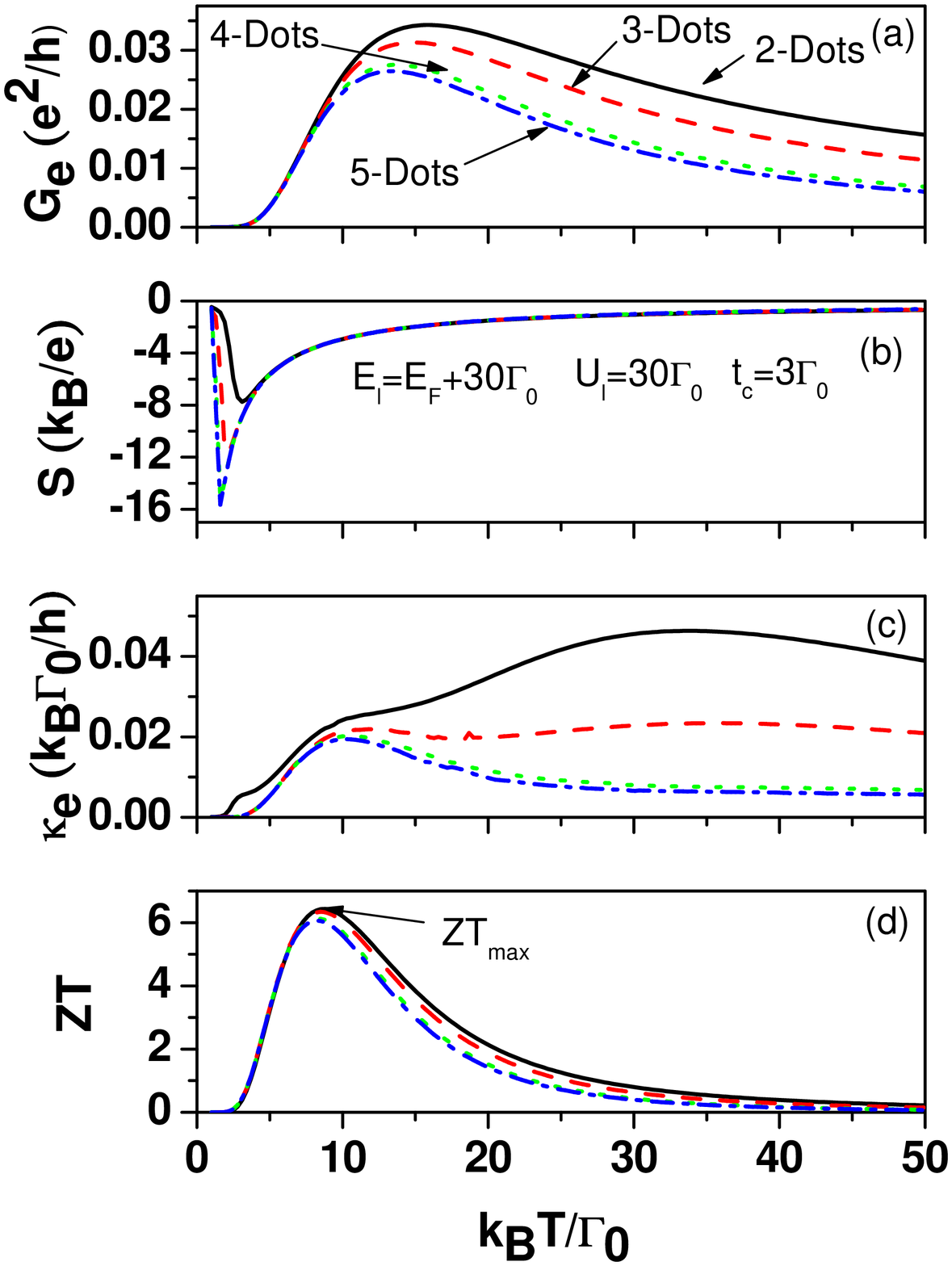}
\caption{$G_e$, $S$, $\kappa_e$, and ZT as functions of temperature
for SQDC with the QD number $N$ varying from 2 to 5. Symmetrical
tunneling rates $\Gamma_L=\Gamma_R=\Gamma_0$ are used.
$E_{\ell}=E_0=E_F+30\Gamma_0$, intradot Coulomb interactions
$U_{\ell}=U_0=30\Gamma_0$, and electron hopping strengths
$t_{\ell,j}=t_c=3\Gamma_0$.$\Gamma_L=\Gamma_R=\Gamma=\Gamma_0$. }
\end{figure}

\begin{figure}[h]
\centering
\includegraphics[scale=0.4]{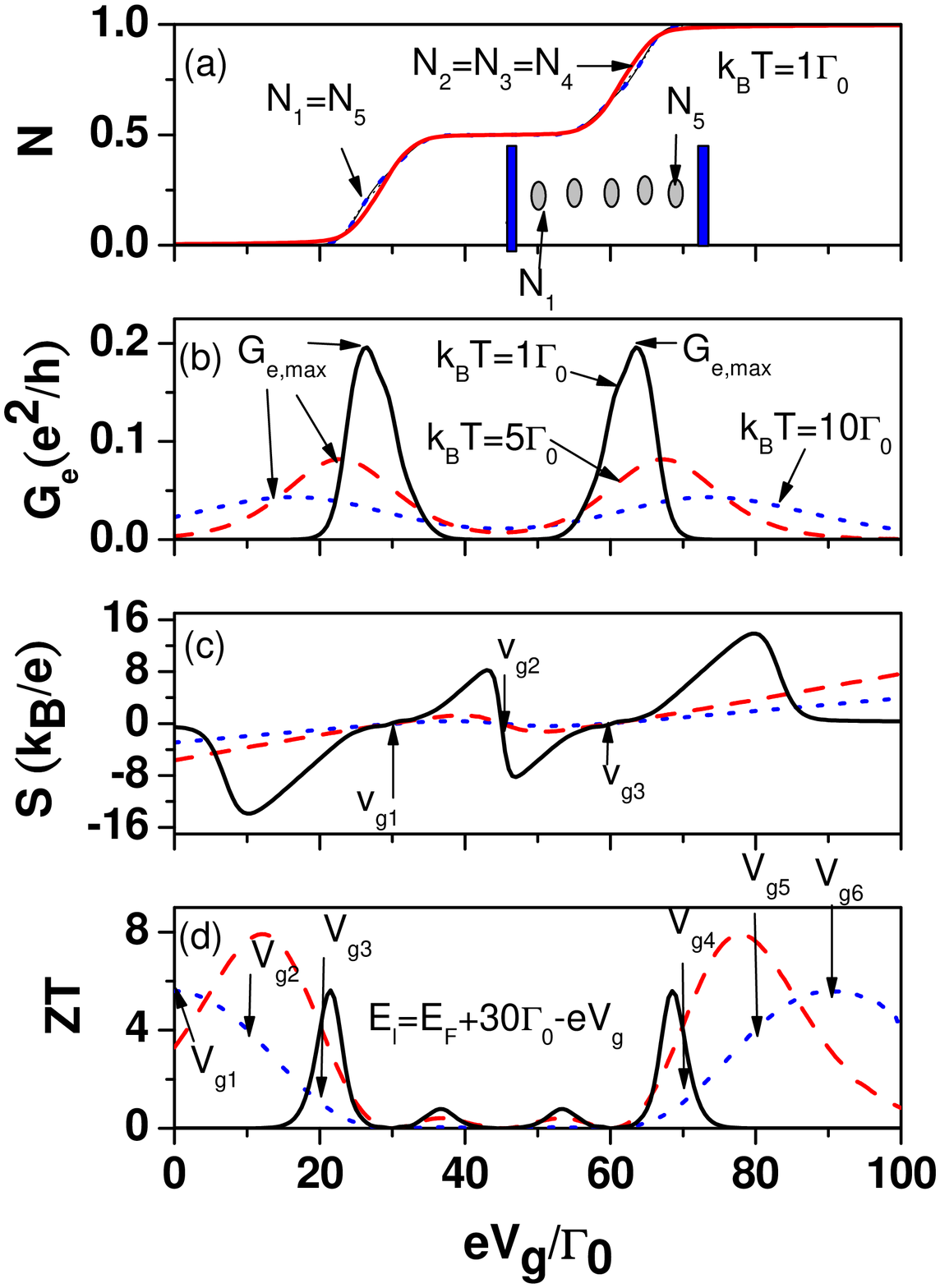}
\caption{Electron occupation number ($N$), electrical conductance
$(G_e)$, Seebeck coefficient ($S$), and ZT as functions of gate
voltage ($V_g$) for various temperatures for N=5.  Other physical
parameters are the same as those of Fig. 8.
 }
\end{figure}

\begin{figure}[h]
\centering
\includegraphics[scale=0.4]{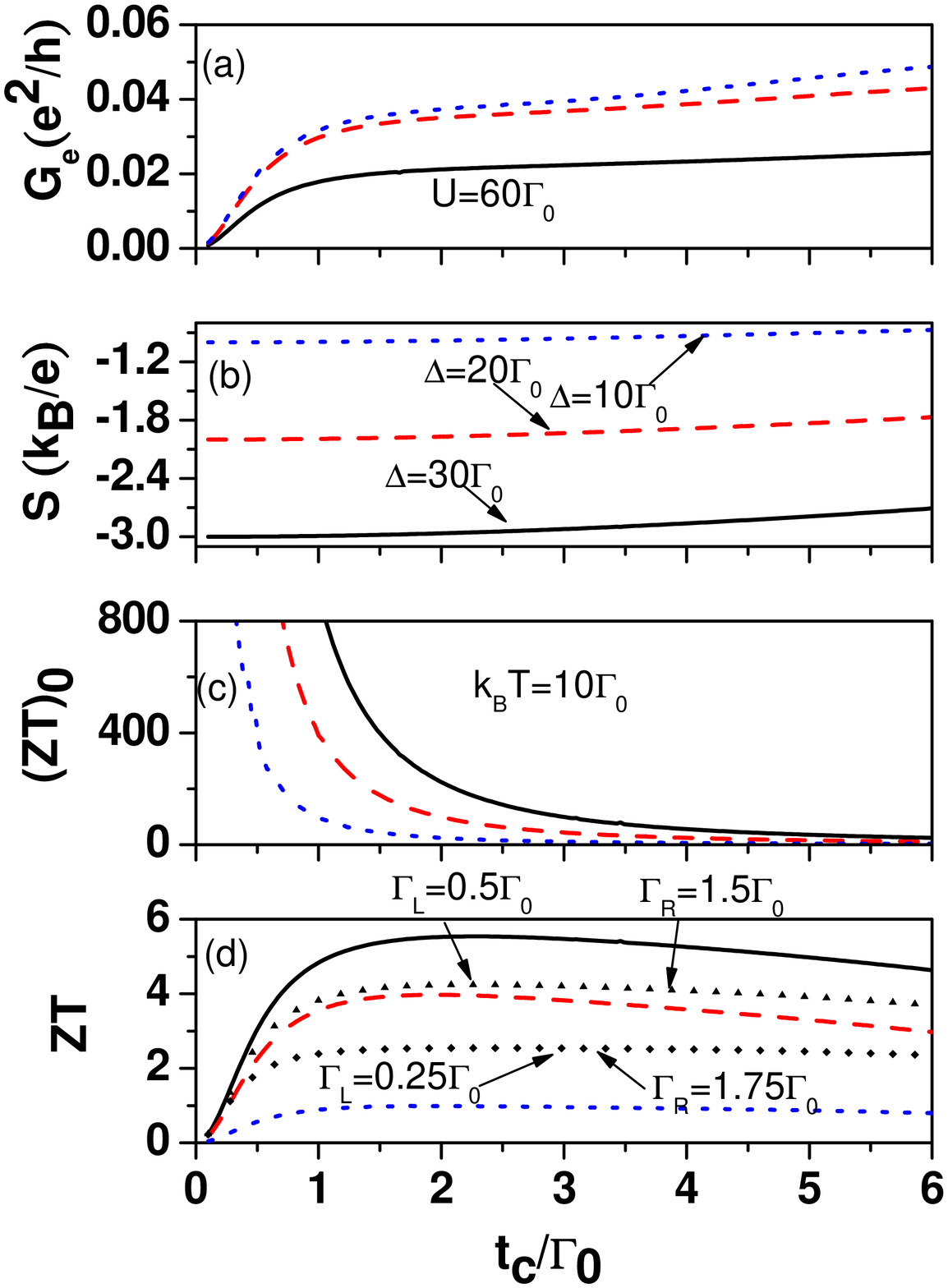}
\caption{ $G_e$, $S$, (ZT)$_0$, and $ZT$ as functions of $t_c$ at
$k_BT=10\Gamma_0$ for various detuning energies. $U=60\Gamma_0$ and
$\Gamma=\Gamma_0$. The curves marked by filled triangles and
diamonds in Fig.~10(d) are for $\Delta=30\Gamma_0$, but with
asymmetric tunneling rates $\Gamma_L \neq \Gamma_R$.}
\end{figure}

\end{document}